\documentclass[12pt]{article}
\usepackage[a4paper, top=16mm, text={170mm, 248mm}, includehead, includefoot, hmarginratio=1:1, heightrounded]{geometry}
\usepackage{amsmath,amssymb,mathrsfs,amsthm,tikz,shuffle,paralist}
\usepackage{dsfont}
\usepackage{color}
\definecolor{darkred}{RGB}{173,34,48}

\usetikzlibrary{snakes}
\usetikzlibrary{calc}
\usetikzlibrary{decorations}

\usepackage[all]{xypic}
\usepackage{jheppub}
\usepackage{hyperref}
\usepackage{subfigure}
\usepackage{caption}

\title{Truncated cluster algebras and Feynman integrals with algebraic letters}

\date{\today}
\author[a,b,c,d,e]{Song He}
\author[a,d]{Zhenjie Li} 
 \author[a,d]{Qinglin Yang}%
\affiliation[a]{CAS Key Laboratory of Theoretical Physics, Institute of Theoretical Physics, Chinese Academy of Sciences, Beijing 100190, China}
\affiliation[b]{
School of Fundamental Physics and Mathematical Sciences, Hangzhou Institute for Advanced Study, UCAS, Hangzhou 310024, China}
\affiliation[c]{ICTP-AP
International Centre for Theoretical Physics Asia-Pacific, Beijing/Hangzhou, China}
\affiliation[d]{School of Physical Sciences, University of Chinese Academy of Sciences, No.19A Yuquan Road, Beijing 100049, China}
\affiliation[e]{Peng Huanwu Center for Fundamental Theory, Hefei, Anhui 230026, P. R. China}
\emailAdd{songhe@itp.ac.cn}
\emailAdd{lizhenjie@itp.ac.cn}
\emailAdd{yangqinglin@itp.ac.cn}

\abstract{We propose that the symbol alphabet for classes of planar, dual-conformal-invariant Feynman integrals can be obtained as truncated cluster algebras purely from their kinematics, which correspond to boundaries of (compactifications of) $G_+(4,n)/T$ for the $n$-particle massless kinematics. For one-, two-, three-mass-easy hexagon kinematics with $n=7,8,9$, we find finite cluster algebras $D_4$, $D_5$ and $D_6$ respectively, in accordance with previous result on alphabets of these integrals. As the main example, we consider hexagon kinematics with two massive corners on opposite sides and find a truncated affine $D_4$ cluster algebra whose polytopal realization is a co-dimension 4 boundary of that of $G_+(4,8)/T$ with 39 facets; the normal vectors for 38 of them correspond to g-vectors and the remaining one gives a limit ray, which yields an alphabet of $38$ rational letters and $5$ algebraic ones with the unique four-mass-box square root. We construct the space of integrable symbols with this alphabet and physical first-entry conditions, whose dimension can be reduced using conditions from a truncated version of cluster adjacency. Already at weight $4$, by imposing last-entry conditions inspired by the $n=8$ double-pentagon integral, we are able to uniquely determine an integrable symbol that gives the algebraic part of the most generic double-pentagon integral. Finally, we locate in the space the $n=8$ double-pentagon ladder integrals up to four loops using differential equations derived from Wilson-loop $d\log$ forms, and we find a remarkable pattern about the appearance of algebraic letters. }
\begin{document}

\maketitle
\section{Introduction and review}
Recent years have witnessed enormous progress in computing and understanding analytic structures of scattering amplitudes in QFT. Not only do these developments greatly pushed the frontier of perturbative calculations relevant for high energy experiments, but they also offer deep insights into the theory itself and exhibit surprising connections with mathematics. An outstanding example is the ${\cal N}=4$ super-Yang-Mills theory (SYM), where one can perform calculations that were unimaginable before and discover rich mathematical structures underlying them. For example, positive Grassmannian~\cite{Arkani-Hamed:2016byb} and the amplituhedron~\cite{Arkani-Hamed:2013jha} have provided a new geometric formulation for its planar integrand to all loop orders.

A related direction which we focus on in this paper concerns the deep connection between singularities of loop amplitudes in planar SYM and cluster algebras related to positive Grassmannians~\cite{Golden:2013xva}. It was first discovered in~\cite{Arkani-Hamed:2016byb} that Grassmannian cluster algebra~\cite{
speyer2005tropical} naturally appear from the quiver dual to plabic graphs that compute loop integrand of the theory. Remarkably, it has been realized in~\cite{Arkani-Hamed:2016byb} that cluster algebras of Grasmmannian $G(4,n)$ are directly relevant for branch cuts of loop amplitudes with $n$ particles. More precisely, the ${\cal A}$  coordinates of $G(4,n)$ cluster algebras are related to symbol~\cite{Goncharov:2010jf, Duhr:2011zq} letters of amplitudes: the $9$ letters of six-particle amplitudes and $42$ letters of seven-particle ones are nicely explained by $A_3$ and $E_6$ cluster algebras, respectively~\footnote{The rank of the cluster algebra is given by the dimension of the kinematic space parameterized by momentum twistors, which is $3(n-5)$ for $G_+(4,n)/T$ due to dual conformal symmetry.}; they have been exploited for bootstrap program to impressively high orders~\cite{Dixon:2011pw,Dixon:2014xca,Dixon:2014iba,Drummond:2014ffa,Dixon:2015iva,Caron-Huot:2016owq,Dixon:2016nkn,Drummond:2018caf, Caron-Huot:2019vjl, Caron-Huot:2019bsq, Dixon:2020cnr, Caron-Huot:2020bkp}. Perhaps even more surprisingly, cluster algebras seem to dictate how different singularities of amplitudes are related to each other, known as ``cluster adjacency"~\cite{Drummond:2017ssj,Drummond:2018caf,Drummond:2018dfd}, which are closely related to the so-called extended Steinmann relations~\cite{Caron-Huot:2016owq,Dixon:2016nkn,Caron-Huot:2019bsq,Caron-Huot:2020bkp}. For $n=6,7$, all known amplitudes exhibit a remarkable pattern that only ${\cal A}$ coordinates that belong to the same cluster can appear adjacent in the symbol.

Beyond $n=6,7$, Grassmannian cluster algebras for $G(4,n)$ for $n\geq 8$ become infinite, thus it is an important question how to truncate it to give a finite symbol alphabet. Moreover, as already seen for one-loop N${}^2$MHV, amplitudes with $n\geq 8$ generally involve letters that cannot be expressed as rational functions of Pl\"ucker coordinates of the kinematics $G(4,n)/T$; more non-trivial algebraic ($=$ non-rational throughout the paper) letters appear in new computations based on ${\bar Q}$ equations~\cite{CaronHuot:2011kk} for two-loop NMHV amplitudes for $n=8$ and $n=9$~\cite{Zhang:2019vnm, He:2020vob}. This means that in addition to the truncation, new ingredients are needed in the context of Grassmanian cluster algebras to explain these and more algebraic letters. A solution to both problems has been proposed using tropical positive Grassmannian~\cite{speyer2005tropical} and related tools for $n=8$~\cite{Drummond:2019qjk, Drummond:2019cxm, Henke:2019hve, Arkani-Hamed:2019rds,Arkani-Hamed:2020cig, Herderschee:2021dez} and very recently for $n=9$~\cite{Henke:2021avn,Ren:2021ztg}~\footnote{In this paper, we consider polytopal realization of $G_+(4,n)/T$ as explicitly computed in~\cite{He:2020ray} using Minkowski sum based on stringy integrals~\cite{Arkani-Hamed:2019mrd}, which is dual to tropical positive Grassmannian. We will not consider tropical Grassmannian explicitly, though the latter can be recovered from our polytope easily. Also see \cite{Cachazo:2019ngv,Cachazo:2019xjx,Drummond:2020kqg,Parisi:2021oql,Lukowski:2020dpn} for recent studies in a different context.}. Another method for explaining the alphabet has been proposed using Yangian invariants or the associated collections of plabic graphs~\cite{Mago:2020kmp, He:2020uhb,Mago:2020nuv,Mago:2021luw}.

On the other hand, ${\cal N}=4$ SYM has been proven to be an extremely fruitful laboratory for the study of Feynman integrals ({\it c.f.} \cite{ArkaniHamed:2010gh,Drummond:2010cz, Caron-Huot:2018dsv, Henn:2018cdp, Bourjaily:2018aeq, Herrmann:2019upk} and references therein). Remarkably, the connection to cluster algebras seems to extend to individual Feynman integrals as well, {\it e.g.} the same $A_3$ and $E_6$ control $n=6,7$ multi-loop integrals in ${\cal N}=4$ SYM~\cite{Caron-Huot:2018dsv, Drummond:2017ssj}. Very recently, cluster algebra structures have been discovered for Feynman integrals beyond those in planar ${\cal N}=4$ SYM~\cite{Chicherin:2020umh}: there is strong evidence for a $C_2$ cluster algebra and adjacency for four-point Feynman integrals with an off-shell leg, and various cluster-algebra alphabets have been found for one-loop integrals, and general five-particle alphabet which play an important role in recent two-loop computations~\cite{Abreu:2018aqd,Chicherin:2018old,Chicherin:2018yne}. Apart from connection to cluster algebras, knowledge of alphabet (and further information) can be used for bootstrapping Feynman integrals~\cite{Chicherin:2017dob, Henn:2018cdp} (see also \cite{Dixon:2020bbt}). In~\cite{He:2021esx}, we identified cluster algebras and certain adjacency conditions for a class of finite, dual conformal invariant (DCI)~\cite{Drummond:2006rz,Drummond:2007aua} Feynman integrals to high loops,  based on recently-proposed Wilson-loop $d\log$ representation~\cite{He:2020uxy} (see~\cite{Bourjaily:2018aeq} for a closely-related Feynman-parameter representation). For ladder integrals with possible ``chiral pentagons" on one or both ends (without any square roots), we find a sequence of cluster algebras $D_2, D_3, \cdots, D_6$  for their alphabets, depending on $n$ and the kinematic configurations. 

Note that some integrals share the same (or almost the same) alphabet, such as $A_3$ or $E_6$, as the amplitudes for $n=6,7$ since the kinematics is just that of the $n$ massless particles; other integrals depend on less kinematic variables, {\it e.g.} double-penta-ladder integrals for $n=7$ (with two legs on a corner) depend on $4$ out of 6 variables and the alphabet turns out to be $D_4$. What is non-trivial about results in~\cite{He:2021esx} is that for such a class of Feynman integrals we always find a cluster algebra, which is a sub-algebra of that of $G(4,n)$ (as opposed to a random subset), which is already interesting for $n=7$ but more so for $n=8,9$ {\it etc.}~\footnote{At one or two loops, we usually only see a subset of the full alphabet, but at high enough loops, the alphabet becomes stable and exactly correspond to {\it e.g.} those type-$D$ cluster algebras.}. Another intriguing observation is that the alphabet and cluster algebra structure of these DCI Feynman integrals seem to be independent of details such as numerators or loop orders, but controlled by the kinematics only. It is natural to ask if one can predict the alphabet and possible adjacency conditions for these DCI integrals, as well as those with algebraic letters, from cluster algebra considerations. 
In this paper, we take a first step in making prediction for the alphabet of DCI Feynman integrals from cluster algebras associated with their kinematics, which correspond to boundaries $G_+(4,n)/T$
.  We propose that for certain kinematics which can be parametrized by a positroid cell of $G_+(4,n)$, the candidate alphabet for Feynman integrals is given by a cluster algebra obtained from an initial quiver which is the dual of the plabic graph; we are done if the resulting cluster algebra is finite (these include the type-$D$ cases in~\cite{He:2021esx}), but generically it becomes infinite just as $G(4,n)$ cluster algebra for $n\geq 8$, and we need truncation. The procedure is equivalent to that in~\cite{Drummond:2019cxm, Henke:2019hve, Arkani-Hamed:2019rds} (see also \cite{He:2020ray}): we construct a polytopal realization for this boundary of $G_+(4,n)/T$ by taking Minkowski sum of Newton polytopes of (non-vanishing) Pl\"ucker coordinates, and the facets of the polytope (dual to the rays of tropical positive Grassmannian) teaches us how to truncate the cluster algebra and possibly include algebraic letters. We will loosely refer to the alphabet that comes from this procedure as a {\it truncated cluster algebra} associated with the kinematics~\footnote{It is important to note that compactification introduced by Minkowski sum/tropicalization always give truncations, even for cases with finite cluster algebra; {\it e.g.} for $G_+(4,7)/T$, there are various different compactifications which all give an alphabet with $42$ letters, but the polytopes/tropical fans are different! In this paper we stick to the analog of Speyer-Williams fan by using Minkowski sum using all (non-vanishing) Pl\"ucker coordinates.}. 
We expect that the truncation using Minkowski sum or tropicalization of all non-vanishing Pl\"ucker coordinates commutes with taking boundaries in $G_+(4,n)/T$, thus alternatively we can just take the truncated cluster algebra of the latter and go to the corresponding boundary. However, the computation for $G_+(4,n)/T$ becomes extremely complicated beyond $n=8$, thus for low-dimensional boundaries it makes no sense to do the full computation and then go down. Our proposal makes it more practical to predict symbol alphabet of higher-point Feynman integrals, especially those with more massive corners (whose kinematics depend on less variables). Moreover, boundaries of $G_+(4,n)/T$ and corresponding truncated cluster algebras deserve investigations on their own({\it c.f.}~\cite{Arkani-Hamed:2020tuz}); a systematic study of these boundaries is beyond the scope of this paper, and we will illustrate our proposal with a few examples which can be applied to classes of Feynman integrals we are interested in. In particular we find a co-dimension 4 boundary of $G_+(4,8)/T$ whose cluster algebra is an affine $D_4$ type. The Minkowski sum gives a polytope with $39$ facets, and we obtain $38$  rational letters plus $5$ algebraic ones.

Another motivation for our study comes from interests in Feynman integrals (and scattering amplitudes) with algebraic letters, which poses certain challenges for multi-loop computations. Using direct integration (either as $d\log$ forms~\cite{He:2020uxy} or in Feynman parametrization form~\cite{Bourjaily:2018aeq}), it is straightforward to evaluate such DCI Feynman integrals to high loops for cases without any square roots. The presence of algebraic letters makes computation difficult and structures obscured due to the need of rationalization and cancellation of spurious square roots~\cite{He:2020lcu, Bourjaily:2019igt}; for example, the symbol of most general double-pentagon integrals contains $16$ square roots of four-mass-box type, and for each of them there are $5$ (multiplicative independent) algebraic letters. The technical difficulty involved is almost identical to that in computing two-loop NMHV amplitudes from ${\bar Q}$ equations~\cite{Zhang:2019vnm,He:2020vob}, and extensions to higher loops become more and more difficult. It is an interesting and difficult problem in computing (the symbol) of these integrals and amplitudes at higher loops, and understanding structures of the result involving algebraic letters.

\begin{figure}[htbp]
    \centering
\begin{tikzpicture}[baseline={([yshift=-.5ex]current bounding box.center)},scale=0.18]
        \draw[black,thick] (0,0)--(0,5)--(4.76,6.55)--(7.69,2.5)--(4.76,-1.55)--cycle;
        \draw[black,thick] (-15,5)--(-19.76,6.55)--(-22.69,2.5)--(-19.76,-1.55)--(-15,0);
        \draw[decorate, decoration=snake, segment length=12pt, segment amplitude=2pt, black,thick] (4.76,6.55)--(4.76,-1.55);
        \draw[decorate, decoration=snake, segment length=12pt, segment amplitude=2pt, black,thick] (-19.76,6.55)--(-19.76,-1.55);
        \draw[black,thick] (9.43,1.5)--(7.69,2.5)--(9.43,3.5);
        \draw[black,thick] (4.76,6.55)--(5.37,8.45);
        \draw[black,thick] (4.76,-1.55)--(5.37,-3.45);
        \draw[black,thick] (0,5)--(-5,5)--(-5,0)--(0,0);
        \draw[black,thick,densely dashed] (-5,5)--(-10,5);
        \draw[black,thick,densely dashed] (-5,0)--(-10,0);
        \draw[black,thick] (-10,0)--(-10,5)--(-15,5)--(-15,0)--cycle;
        \draw[black,thick] (-19.76,6.55)--(-20.37,8.45);
        \draw[black,thick] (-19.76,-1.55)--(-20.37,-3.45);
        \draw[black,thick] (-24.69,3.5)--(-22.69,2.5)--(-24.69,1.5);
        \filldraw[black] (-20.37,8.45) node[anchor=south] {{8}};
        \filldraw[black] (5.37,8.45) node[anchor=south] {{1}};
        \filldraw[black] (9.43,3.5) node[anchor=west] {{2}};
        \filldraw[black] (9.43,1.5) node[anchor=west] {{3}};
        \filldraw[black] (5.37,-3.45) node[anchor=north] {{4}};
        \filldraw[black] (-20.37,-3.45) node[anchor=north] {{5}};
        \filldraw[black] (-24.69,1.5) node[anchor=east] {{6}};
        \filldraw[black] (-24.69,3.5) node[anchor=east] {{7}};
    \end{tikzpicture}\quad \quad \begin{tikzpicture}[baseline={([yshift=-.5ex]current bounding box.center)},scale=0.18]
        \draw[black,thick] (0,0)--(4,0)--(6,3.46)--(4,6.93)--(0,6.93)--(-2,3.4)--cycle;
       \draw[black,thick] (4,6.93)--(5,8.66);
        \draw[black,thick] (4,0)--(5,-1.73);
        \filldraw[black] (5,-1.73) node[anchor=north west] {{$4$}};
        \filldraw[black] (7.73, 2.26) node[anchor=north west] {{$3$}};
        \filldraw[black] (7.73, 4.66) node[anchor=south west] {{$2$}};
        \filldraw[black] (5,8.66) node[anchor=south west] {{$1$}};
        \filldraw[black] (-1,8.66) node[anchor=south east] {{$8$}};
        \filldraw[black] (-3.73,2.46) node[anchor=north east] {{$6$}};
        \filldraw[black] (-3.73,4.46) node[anchor=south east] {{$7$}};
        \filldraw[black] (-1,-1.73) node[anchor=north east] {{$5$}};
        \draw[black,thick] (7.73,2.46)--(6,3.46)--(7.73,4.46);
        \draw[black,thick] (-3.73,2.46)--(-2,3.46)--(-3.73,4.46);
        \draw[black,thick] (0,0)--(-1,-1.73);
        \draw[black,thick] (0,6.93)--(-1,8.66);
    \end{tikzpicture}   
\end{figure}
Among these integrals we consider, arguably the simplest all-loop series involving non-trivial algebraic letters is the class of double-penta-ladder integrals $\Omega_L(1,4,5,8)$
\cite{Bourjaily:2018aeq,He:2020uxy} (as shown in the figure above); the kinematics involved can be drawn as a hexagon with two massive corners on the opposite sides, which corresponds to the co-dimension $4$ boundary of $G_+(4,8)/T$. As we will explain, the $L$-loop integral can be written as two-fold integral of $(L-1)$-loop on (alternatively a pair of nice first-order differential operators reduce the former to the latter). However, unlike the seven-point counterpart (or those higher-point cases without square roots), performing such integrations become tricky due to the presence of square roots, which also prevents us for seeing underlying structures concerning algebraic letters. Now equipped with the alphabet from truncated cluster algebra (and physical discontinuity conditions), we can construct the space of all possible multiple polylogarithm functions (MPL) at symbol level  which can be further reduced by adjacency conditions, and we  conjecture that the space includes all DCI integrals with this kinematics. This ``bootstrap” strategy can be viewed as an extension of results in~\cite{Henn:2018cdp} to include algebraic letters~\footnote{This is in spirit a bit different from bootstrapping amplitudes/form factors since in principle we have Wilson-loop $d\log$ forms/differential equations which determine the answer; in some sense all we need to do is to locate the solution.}. Already at weight $4$, we find that simply by imposing last entry conditions implied by $d\log$ form or differential equations, the part containing square root is uniquely determined! Moreover, this weight $4$ function with algebraic letters turns out to be the ``seed" for (the algebraic part of) the most general $n=12$ double-pentagon integrals: the latter can be obtained by the sum of $16$ functions with relabeling; this suggests that the $n=12$ case contains $16$ such truncated cluster algebras. 

Moving to higher weights, we can easily determine $\Omega_L(1,4,5,8)$ to four loops (as a strong support for the alphabet and adjacency conditions), by imposing differential equations and boundary conditions obtained from $d\log$ forms, which circumvent the need of rationalization at all. Furthermore, we discover some nice pattern about the algebraic letters, which confirms a conjecture we had for these integrals: at least through four loops, non-trivial algebraic letters only appear on the third entry of the symbol, with the accompanying first two entries being that of the four-mass box! Thus for the algebraic part, the highly non-trivial procedure of performing $d\log$ integrals/solving differential equations amounts to simply ``translating" the first three entries, and ``attaching" rational letters in subsequent entries. Similar observations have been made for $k+\ell=3$ level of $n=8$ amplitudes~\cite{Zhang:2019vnm}, and we hope our results can provide a starting point for future investigation into similar structures of multi-loop integrals and amplitudes with algebraic letters.

The rest of the paper is organized as follows. We first give a quick review of cluster algebras and polytopes from certain stringy integrals, which will be used in our study of truncated cluster algebras. In sec.~\ref{sec:2}, we describe our procedure for predicting alphabet of Feynman integrals: after presenting warm-up examples for finite cases $D_4, D_5$ and $D_6$, we give the truncated affine $D_4$ as the alphabet for ``two-mass-opposite" hexagon kinematics. It consists of $38$ rational letters associated with $g$-vectors, and $5$ algebraic ones associated with the limit ray (containing the unique four-mass-box square root). In sec.~\ref{sec:3}, we move to constructing the cluster function space at symbol level and determine $\Omega_L(1,4,5,8)$ inside the space. With first entry conditions and constraints from a truncated version of cluster adjacency, we obtain the reduced space up to weight $6$, and already at weight $4$ one can determine a unique function responsible for the algebraic part of most generic double-pentagon integrals. We then discuss constraints for $\Omega_L(1,4,5,8)$ including last entries, differential equations {\it etc.}. Finally, we determine $\Omega_L(1,4,5,8)$ up to four loops and discuss the pattern concerning the algebraic letters. 

\subsection{Review of cluster algebras and polytopes from stringy integrals}

Let us first give a lightening review of cluster algebras~\cite{fomin2002cluster,fomin2003cluster,berenstein2005cluster,fomin2007cluster}, where we only give necessary ingredients needed in this paper.  Cluster algebras are commutative algebras with a particular set of generators ${\cal A}_i$, known as the cluster ${\cal A}$-coordinates; they are grouped into {\it clusters} which are subsets of rank $n$. From an initial cluster, one can construct all the ${\cal A}$-coordinates by {\it mutations} acting on ${\cal A}$'s (the so-called frozen coordinates or coefficients can also be included, which do not mutate). 

Cluster variables in a cluster are related by arrows, which forms a quiver $Q$ (without 2-cycles, i.e. arrows $*\to \cdot\to *$). Then we associate $Q$ with an skew-symmetric exchange matrix $B(Q) = (b_{ij})$ by $b_{ij} = -b_{ji} = l$ whenever there are $l$ arrows from vertex $i$ to vertex $j$. 
Suppose we mutate the vertex $k$, then the exchange matrix of the mutated quiver reads
\[
b_{i j}^{\prime}=\left\{\begin{array}{ll}
-b_{i j} & \text { if } i=k \text { or } j=k \\
b_{i j}+\operatorname{sgn}\left(b_{i k}\right)\left[b_{i k} b_{k j}\right]_{+} & \text {otherwise, }
\end{array}\right.
\]
where $[x]_+:=\max(x,0)$, and the cluster variable $x_k$ on vertex $k$ is mutated to $x_k'$ given by
\[
x_{k}^{\prime}x_{k}=\prod_{i=1}^{n} x_{i}^{\left[b_{i k}\right]_{+}}+ \prod_{i=1}^{n} x_{i}^{\left[-b_{i k}\right]_{+}}.
\]
In general, mutations generate infinite number of cluster variables. As classified in \cite{fomin2003cluster},  only cluster algebras whose quiver can be mutated from a Dynkin diagram of type $A,B,C,D,E,F,G$ have finite number of cluster variables, known as the cluster algebra of finite type. The number of cluster variables (dimension) $N$ for finite types read:
\begin{align*}
N(A_n)=n(n+3)/2,\ N(B_n)=N(C_n)=n(n+1),\ N(D_n)=n^2,\\
N(E_6)=42,\ N(E_7)=70,\ N(E_8)=128,\ N(F_4)=28,\ N(G_2)=8.
\end{align*}

According to \cite{fomin2007cluster}, one can further assign a coefficient to a vertex, where the coefficient should be a monomial of some given free variables
. Then the mutation rule for cluster variable $x_k$ and the coefficient $y_k$ on vertex $k$ reads
\begin{equation}
y_{j}^{\prime}=\left\{\begin{array}{ll}
y_{k}^{-1} & \text { if } j=k, \\
y_{j} y_{k}^{[b_{k j}]_{+}}\left(y_{k} \oplus 1\right)^{-b_{k j}} & \text { if } j \neq k,
\end{array}\right.
\end{equation}
and 
\begin{equation}
x_{k}^{\prime}x_{k}=\frac{y_{k}}{y_{k} \oplus 1} \prod_{i=1}^{n} x_{i}^{\left[b_{i k}\right]_{+}}+\frac{1}{y_{k} \oplus 1} \prod_{i=1}^{n} x_{i}^{\left[-b_{i k}\right]_{+}},
\end{equation}
where the addition $\oplus$ for monomials of free variables $\{u_i\}$ is defined by 
\[
\prod_{j} u_{j}^{a_{j}} \oplus \prod_{j} u_{j}^{b_{j}}=\prod_{j} u_{j}^{\min \left(a_{j}. b_{j}\right)}.
\]
If the coefficients of a cluster are exactly the chosen free variables, then we call that this cluster has principal coefficients. Therefore, starting form a cluster $\{x_i\}_{i=1,\dots,n}$ with exchange matrix $B=(b_{ij})$ and principal coefficients $\{y_i\}_{i=1,\dots,n}$, the cluster variable on vertex $k$ after a series of mutations of vertices $\mathbf{v}$ is a rational function of initial cluster variables and coefficients
\[
    X_{\mathbf{v},k}=X_{\mathbf{v},k}(x_1,\dots,x_n;y_1,\dots,y_n).
\]
Furthermore, if one defines a $\mathbb{Z}^n$-grading on $\mathbb{Z}[x_1^{\pm 1},\dots,x_n^{\pm 1};y_1,\dots,y_n]$ by $\deg(x_i)=\mathbf{e}_i$ ($1$ in the $i$-th component and $0$ in the rest) and $\deg(y_j)=-\sum_ib_{ij}\mathbf{e}_i$, then $X_{\mathbf{v},k}$ is homogeneous, and its degree $g_{\mathbf{v},k}=(g_{\mathbf{v},k}^1,\dots, g_{\mathbf{v},k}^n)\in \mathbb{Z}^n$ is called its $g$-vector. Another useful polynomial related to $X_{\mathbf{v},k}$ is the $F$-polynomial
\begin{equation}
F_{\mathbf{v},k}(y_1,\dots,y_n):= X_{\mathbf{v},k}(1,\dots,1;y_1,\dots,y_n).
\end{equation}
Once known the $F$-polynomial and $g$-vector, we can recover the whole $X_{\mathbf{v},k}$ by
\begin{equation}
    X_{\mathbf{v},k}(x_1,\dots,x_n;y_1,\dots,y_n)=x_1^{g_{\mathbf{v},k}^1}\cdots x_n^{g_{\mathbf{v},k}^n} F_{\mathbf{v},k}\bigg(y_{1} \prod_i x_{i}^{b_{i 1}},\dots,y_{n} \prod_i x_{i}^{b_{i n}}\bigg).
\end{equation}
There is even a conjecture \cite{fomin2007cluster} to read $g$-vector from $F$-polynomial alone: If $F_{\mathbf{v},k}\neq 1$, then
\begin{equation}
y_1^{g_{\mathbf{v},k}^1}\cdots y_n^{g_{\mathbf{v},k}^n}=\frac{F_{\mathbf{v},k}|_{\text{Trop}}(y_{1}^{-1},\dots,y_{n}^{-1})}{F_{\mathbf{v},k}|_{\text{Trop}}\big(\prod_i y_{i}^{b_{i 1}},\dots,\prod_i y_{i}^{b_{i n}}\big)},
\end{equation}
where $F_{\mathbf{v},k}|_{\text{Trop}}$ means that $+$ is replaced by $\oplus$ in the $F$-polynomial.

Associated with a finite-type cluster algebra (or more generally a truncated cluster algebra), there is a natural space of polylogarithm functions, whose alphabet is given cluster variables (equivalently they can be chosen as $N{-}n $ $F$-polynomials and the $n$ principle coefficients). A cluster function $I^{(w)}$ \cite{Golden:2014xqa,Parker:2015cia} of transcendental weight $w$ is defined such that its differential has the form

\begin{equation}
    d I^{(w)}=\sum_i I_i^{(w-1)} d \log X_i
\end{equation}
where $I^{(w-1)}_i$ are cluster functions of transcendental weight $w-1$ and $X_i$ are cluster ${\cal A}$-coordinates (or $F$- polynomials). We see that the alphabet of a cluster function is by definition the corresponding cluster algebra.

Already for finite-type cluster algebras, it is natural to consider the so-called {\it cluster string integrals} which are ``stringy canonical forms''~\cite{Arkani-Hamed:2019mrd} associated with cluster polytopes (they also give natural ``cluster configuration spaces''~\cite{Arkani-Hamed:2020tuz} which will not be discussed here). For a finite-type (denoted as $\Phi_n$) cluster algebra with principle coefficients ${\bf f}=(f_1\cdots f_n)$ and $F$-polynomials $F_I({\bf f})$ for $I=n+1\cdots N$~\footnote{From here we will denote principle coefficients using $f_I\equiv F_I$ for $I=1,\cdots, n$, which in our subsequent studies can be chosen to be {\it face variables} of a plabic graph.}, we define:
\begin{equation}
    \mathcal{I}_{\Phi_n}(\{s\})=(\alpha')^n \int_{\mathbb{R}_{>0}^n} \prod_{i=1}^n d\log f_i \prod_{I=1}^N F_I({\bf f})^{\alpha' s_I}\,.
\end{equation}
As $\alpha^\prime\to 0$, leading order of the integral computes the canonical function of cluster polytope of type $\Phi_n$, which is nicely given by the Minkowski sum of the Newton polytopes of the $F$-polynomials. Vertices of this polytope correspond to clusters: whenever two vertices are connected by an edge, one can mutate from one to the other in the cluster algebra. Furthermore, $N(\Phi_n)$ facets of the polytopes correspond to cluster variables $X_{{\bf v},k}(x_1,\cdots,x_n;f_1,\cdots f_n)$, where we can compute outward normal vectors of these facets~\cite{Arkani-Hamed:2019mrd,Li:2020cve} in terms of the exponents of ${\bf f}=\{f_1, \cdots, f_n\}$, {\it i.e.} $\{s_1, \cdots, s_n\}$. Very nicely, these vectors are nothing but the corresponding $g$-vectors. 
Note that these $g$-vectors for cluster variables in any cluster give a cone: the cones for different clusters are non-overlapping, and the union of all cones (known as the cluster fan) covers the full space in any finite type.  

As mentioned, Grassmannian cluster algebras for $G(4,6)$ and $G(4,7)$ are $A_3$ and $E_6$ respectively, and starting at $n=8$ they become infinite. A natural way for truncating an infinite cluster algebra to be finite has been proposed in~\cite{He:2020ray} using a similar {\it Grassmanian string integrals}. With a  positive parametrization of $G_+(k,n)/T$, we can write the integral where the positive polynomials are instead given by all (or a reasonable subset of) the Pl\"ucker coordinates of $G_+(k,n)$. The leading order as $\alpha' \to 0$ is given by the Minkowski sum of Newton polytopes of these polynomials, and one obtains a polytope for the compactification of $G_+(k,n)/T$~\footnote{For $k=2$, this is the well-known Deligne-Mumford compactification~\cite{deligne1969irreducibility,devadoss1999tessellations,Arkani-Hamed:2017mur} for the moduli space ${\cal M}_{0,n}^+$, which gives the $(n-3)$-dimensional associahedron. For $(k,n)=(3,6), (3,7), (3,8)$ we have polytopes that are related to $D_4, E_6, E_8$ cluster algebras, respectively.}. For infinite type, {\it e.g.} $G(4,n)$ with $n\geq8$ (or $G(3,n)$ with $n\geq 9$), by taking the normal vectors for facets of the polytope, we truncate the infinite cluster algebra by identifying a finite set of $g$-vectors, and there will also be some normal vectors which are not $g$-vectors (called exceptional rays). We remark that the truncation is not unique since it depends on the choice of Pl\"ucker coordinates, and it is equivalent to tropical Grassmannian method since the normal vectors are exactly the rays when choosing the same set of Pl\"ucker coordinates for tropicalization~\cite{Drummond:2019qjk,Drummond:2019cxm,Henke:2019hve}. For $G(4,8)$, if we choose the polynomials to be all Pl\"ucker coordinates, the Minkowski sum gives a polytope with $360$ facets, where $356$ normal vectors are $g$-vectors of $G(4,8)$ cluster variables,  and the remaining $4$ are exceptional rays; if we only keep those of the form $\langle ii{+}1jj{+1}\rangle$ and $\langle i{-}1ii{+}1j\rangle$ (which respect parity), we get a $274$-facet polytope, where $272$ are $g$-vectors and the other $2$ are exceptional ones. Moreover, as we will see shortly, at least for $G(4,8)$ case, an exceptional ray turns out to be a {\it limit ray} which naturally give algebraic letters associated with a square root, in addition to those rational ones corresponding to $g$-vectors.

\section{Truncated cluster algebras for Feynman integrals}\label{sec:2}

In this section, we propose an algorithm which predicts symbol alphabet for classes of DCI Feynman integrals with same kinematics. Here the kinematics simply mean the $m$ dual points which the class of Feynman integrals universally depend on, without referring to actual propagator structure or possible numerators. We will refer to such a kinematical configuration as an $m$-gon with certain massless and massive corners, where for each massless (massive) corner, we put one (two) massless legs, with $n$ legs in total for $n\geq m$~\footnote{We can trivially add more than two legs at a massive corner, which gives higher-point Feynman integrals with the same kinematics, thus the $n$ here is the minimal number of legs.}; when $n=m$, all dual points are null separated, which is the kinematics of $n$ massless legs. For example, all off-shell four-point integrals relevant for four-point CFT correlators share the kinematics of an $n=8$ square ($m=4$, with all four corners massive), and in particular all-loop box ladder integrals belong to this class. The $n=7,8$ pentagon-box ladder proposed in~\cite{Drummond:2010cz} belong to $n=7,8$ pentagon ($m=5$) with two or three massive corners, respectively. It is fun to count the dimension of such kinematics with DCI: for each dual point we have 4 degree of freedoms, but when two points are null separated the degree of freedom is reduced by one, and DCI means subtracting $15$ in the end. For two- or three-mass pentagon, the dimension is $4\times 5-3(2)-15=2(3)$ as expected; for four-mass square it is trickier: the kinematics is so special that one of the $15$ redundancies no longer exists, thus we have $4\times 4-14=2$ dimensions as expected~\footnote{We thank Nima Arkani-Hamed for first explaining this to us.}.

For $m=n=6,7$, the symbol alphabet of the amplitude and all DCI integrals computed so far is dictated by the kinematics, which is given by $A_3$ and $E_6$ respectively. What we propose here is a natural extension to more general kinematics with $m<n$, where we identify it as certain boundaries of $G_+(4,n)/T$. This first gives an equivalent way of counting: from $G_+(4,n)/T$ which has dimension $3(n-5)$, generically for each massive corner we go down in dimension by $2$. It is generally unclear how to identify which boundary of $G_+(4,n)/T$ corresponds to a given kinematics, and to find a truncated cluster algebra for its symbol alphabet. In this paper we focus on special cases where the boundary can be identified with a positroid cell of $G_+(4,n)$ (mod torus action)~\cite{Bourjaily:2018aeq}, which can be labelled by plabic graphs.

The algorithm we propose consists of the following steps, which crucially depends on the fact that the kinematics is associated with a positroid cell.

\begin{itemize}

\item By imposing conditions on Pl\"ucker coordinates of $n$ momentum twistors according to the kinematics, we identify a positroid cell $\Gamma$ of $G_+(4,n)$ represented by a plabic graph $G_\Gamma$, which gives a positive parametrization ${\bf Z}_\Gamma$ of the kinematics (after modding out torus action). More precisely, ${\bf Z}_\Gamma(\{f\})$ depends on internal face variables $f_i$ for $i=1,2,\cdots, d$ where $d$ is the dimension of $\Gamma/T$ (we set all but one external face variables to $1$).

\item We define the cluster algebra ${\cal A}_\Gamma$ by applying mutations from the initial quiver diagram, which is the dual of the plabic graph. We use the face variables as principle coefficients which parametrize the positive part of the cluster variety, and we are interested in the $F$-polynomials. We obtain a finite alphabet if the cluster algebra is a finite type. 

\item We consider all non-vanishing Pl\"ucker coordinates (or a subset of them) of ${\bf Z}_\Gamma$, which are positive polynomials of $f$'s (a subset of $F$-polynomials); we take the Minkowski sum of their Newton polytopes, which gives a polytope denoted as ${\cal P}_\Gamma$. We conjecture that ${\cal P}_\Gamma$ is always a boundary of the polytopal realization of $G_+(4,n)/T$ (which is dual to tropical $G_+(4,n)$). 

\item At least a subset of normal vectors for facets of the polytope ${\cal P}_\Gamma$ should coincide with certain $g$-vectors of ${\cal A}_\Gamma$, and the rational alphabet consists of these $F$-polynomials which are associated with these facets (as well as $f_1, \cdots, f_d$). For those exceptional normal vectors that do not correspond to $g$-vectors, we conjecture that they are associated with non-rational letters {\it etc.} which need to be treated differently. 

\end{itemize}

\subsection{Warm-up examples: truncated $D_4$, $D_5$ and $D_6$ cluster algebras}

Let us begin with warm-up examples for one-, two- and three-mass easy hexagon kinematics with $n=7,8,9$. We will not give details of the computation for these finite-type cases, and simply list the positroid cells given in~\cite{Bourjaily:2018aeq} (with decorated permutations and plabic graphs), positive parametrizations of the kinematics, the polytopes from Minkowski sum and the resulting cluster algebras.  
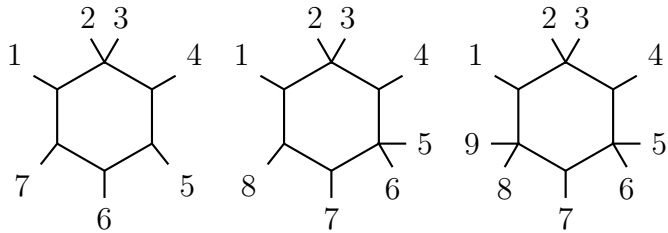
\begin{figure}[htbp]
    \centering
\begin{tikzpicture}[baseline={([yshift=-.5ex]current bounding box.center)},scale=0.18]
        \draw[black,thick] (0,0)--(0,4)--(3.46,6)--(6.93,4)--(6.93,0)--(3.46,-2)--cycle;
        \draw[black,thick] (6.93,4)--(8.66,5);
        \draw[black,thick] (0,4)--(-1.73,5);
        \draw[black,thick] (3.46,-2)--(3.46,-4);
        \filldraw[black] (-1.73,5) node[anchor=south east] {{$1$}};
        \filldraw[black] (2.26,7.73) node[anchor=south] {{$2$}};
        \filldraw[black] (4.66,7.73) node[anchor=south] {{$3$}};
        \filldraw[black] (8.66,5) node[anchor=south west] {{$4$}};
        \filldraw[black] (8.13,-1.53) node[anchor=north west] {{$5$}};
        \filldraw[black] (3.46,-4) node[anchor=north] {{$6$}};
        \filldraw[black] (-1.2,-1.53) node[anchor=north east] {{$7$}};
        \draw[black,thick] (2.46,7.73)--(3.46,6)--(4.46,7.73);
        \draw[black,thick] (0,0)--(-1.2,-1.53);
        \draw[black,thick] (6.93,0)--(8.13,-1.53);
    \end{tikzpicture}
    \begin{tikzpicture}[baseline={([yshift=-.5ex]current bounding box.center)},scale=0.18]
        \draw[black,thick] (0,0)--(0,4)--(3.46,6)--(6.93,4)--(6.93,0)--(3.46,-2)--cycle;
        \draw[black,thick] (6.93,4)--(8.66,5);
        \draw[black,thick] (0,4)--(-1.73,5);
        \draw[black,thick] (3.46,-2)--(3.46,-4);
        \filldraw[black] (-1.73,5) node[anchor=south east] {{$1$}};
        \filldraw[black] (2.26,7.73) node[anchor=south] {{$2$}};
        \filldraw[black] (4.66,7.73) node[anchor=south] {{$3$}};
        \filldraw[black] (8.66,5) node[anchor=south west] {{$4$}};
        \filldraw[black] (8.93,0) node[anchor=west] {{$5$}};
        \filldraw[black] (7.93,-1.73) node[anchor=north] {{$6$}};
        \filldraw[black] (3.46,-4) node[anchor=north] {{$7$}};
        \filldraw[black] (-1.2,-1.53) node[anchor=north east] {{$8$}};
        \draw[black,thick] (2.46,7.73)--(3.46,6)--(4.46,7.73);
        \draw[black,thick] (0,0)--(-1.2,-1.53);
        \draw[black,thick] (8.93,0)--(6.93,0)--(7.93,-1.73);
    \end{tikzpicture}
     \begin{tikzpicture}[baseline={([yshift=-.5ex]current bounding box.center)},scale=0.18]
        \draw[black,thick] (0,0)--(0,4)--(3.46,6)--(6.93,4)--(6.93,0)--(3.46,-2)--cycle;
        \draw[black,thick] (6.93,4)--(8.66,5);
        \draw[black,thick] (0,4)--(-1.73,5);
        \draw[black,thick] (3.46,-2)--(3.46,-4);
        \filldraw[black] (-1.73,5) node[anchor=south east] {{$1$}};
        \filldraw[black] (2.26,7.73) node[anchor=south] {{$2$}};
        \filldraw[black] (4.66,7.73) node[anchor=south] {{$3$}};
        \filldraw[black] (8.66,5) node[anchor=south west] {{$4$}};
        \filldraw[black] (8.93,0) node[anchor=west] {{$5$}};
        \filldraw[black] (7.93,-1.73) node[anchor=north] {{$6$}};
        \filldraw[black] (3.46,-4) node[anchor=north] {{$7$}};
        \filldraw[black] (-1,-1.73) node[anchor=north] {{$8$}};
        \filldraw[black] (-2,0) node[anchor=east] {{$9$}};
        \draw[black,thick] (2.46,7.73)--(3.46,6)--(4.46,7.73);
        \draw[black,thick] (-2,0)--(0,0)--(-1,-1.73);
        \draw[black,thick] (8.93,0)--(6.93,0)--(7.93,-1.73);
    \end{tikzpicture}   
    \caption{One-, two-, three-mass-easy hexagon kinematics with $n=7,8,9$ legs}\label{fig3}
\end{figure}

Let's first consider one-mass kinematics with dual points $(x_1, x_2, x_4, x_5, x_6, x_7)$, which should correspond to a co-dimension 2 positroid. As explained in~\cite{Bourjaily:2018aeq}, the latter can be specified by $\langle n 1 2 3 \rangle =\langle 2 3 4 5\rangle=0$, which gives a decorated permutation $\sigma=\{6, 5, 7, 8, 9, 11, 10\}$, and we find plabic graph
\begin{center}
\includegraphics[scale=0.2]{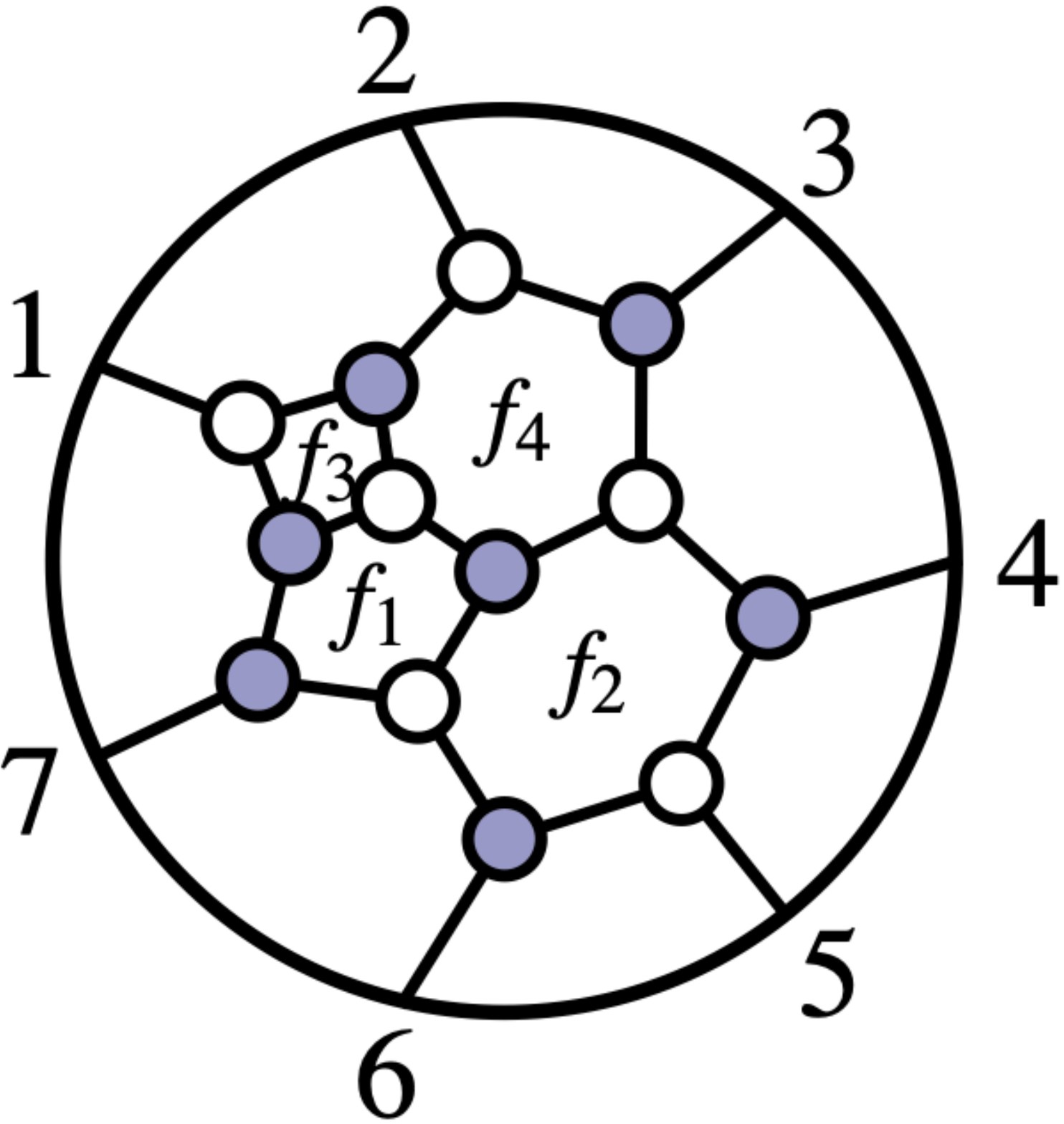}\quad \quad 
\raisebox{.4\height}{\includegraphics[scale=0.45]{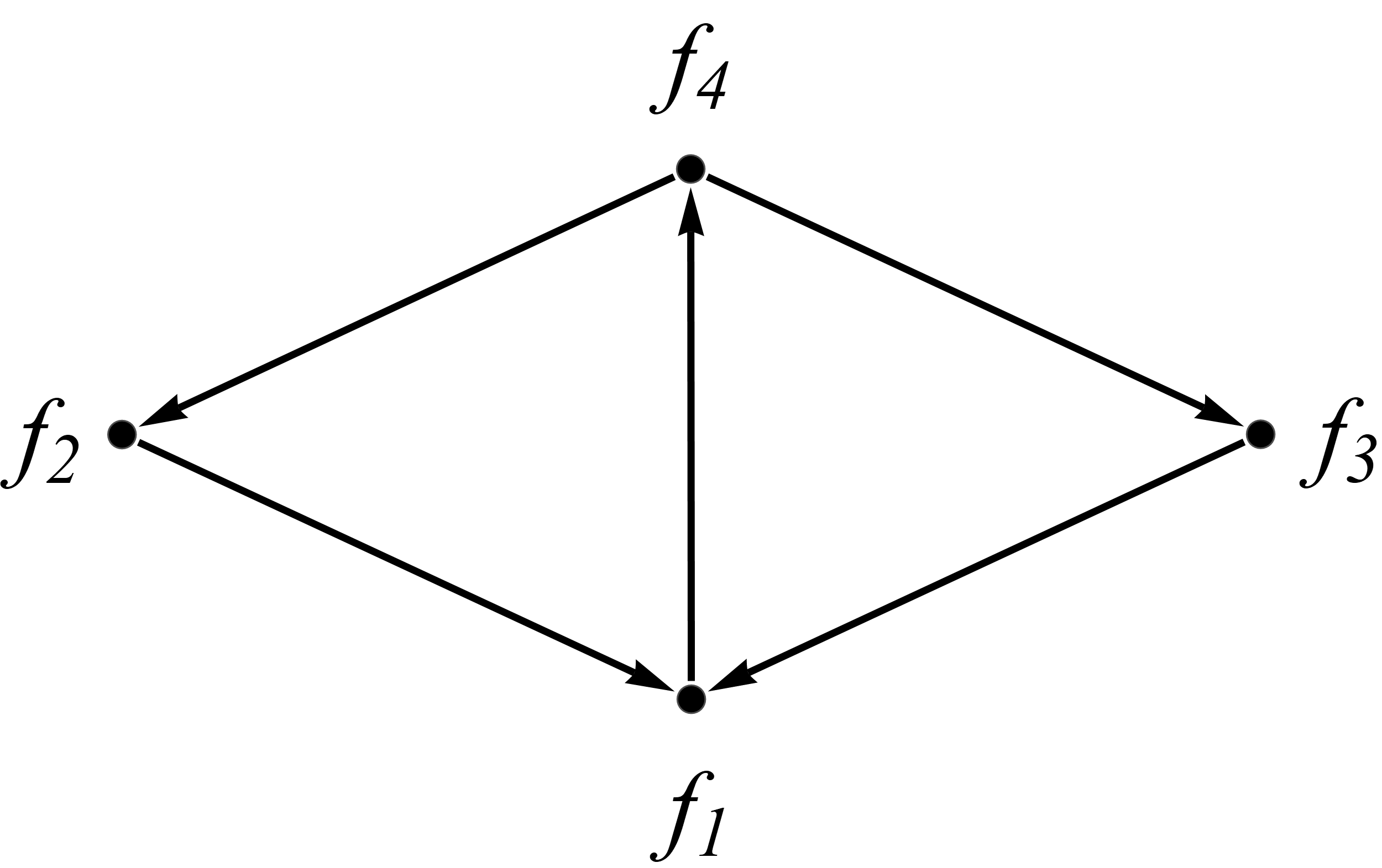}}
\end{center}
For modding out torus action, we fix all but one external face variables to be unity, and the resulting ${\bf Z}$ matrix which positively parametrize the kinematics reads
$$
\footnotesize{\left(
\begin{array}{ccccccc}
 f_3 f_4 & \left(1+f_3\right) f_4 & 1+f_4+f_3 f_4 & 1 & 0 & 0 & 0 \\
 0 & f_1 f_2 f_4 & f_2 \left(1+f_1+f_1 f_4\right) & 1+ f_2+f_1 f_2 & 1 & 0 & 0 \\
 0 & 0 & f_2 & 1+f_2 & 1 & 0 & -1 \\
 0 & 0 & 0 & 1 & 1 & 1 & 0 \\
\end{array}
\right).}
$$

We have drawn the dual quiver diagram of the plabic graph, where we ignored all external facets, on the right. It is easy to see that this is a quiver for the $D_4$ cluster algebra, and as mentioned above, the face variables correspond to principle coefficients assigned to each node. By applying mutation rules we find $16$ cluster variables, which can be identified with $f_1, f_2, f_3, f_4$ and $12$ $F$-polynomials of $f$'s:
\begin{align}
&\{1+f_1,1+f_2,1+f_3,1+f_4, 1+f_2+f_1 f_2,1+f_3+f_1 f_3,1+f_1+f_1 f_4,1+f_4+f_2 f_4,\nonumber\\
&1+f_4+f_3 f_4,1+f_2+ f_3+f_2 f_3+f_1 f_2 f_3,1+f_4+ f_2 f_4+f_3 f_4+f_2 f_3 f_4+f_1 f_2 f_3 f_4\}
\end{align}
Since this is a finite type we have a finite alphabet: we conjecture that any DCI integral with this one-mass hexagon kinematics has a symbol alphabet of the $16$ letters. 
 
On the other hand, by computing all non-vanishing Pl\"ucker coordinates of the ${\bf Z}$ matrix, we find $15$ positive polynomials, which actually already contain $15$ of the above alphabet, except for $1+f_4+f_2 f_4+f_3 f_4+f_2 f_3 f_4$.
Now, we compute the Minkowski sum of Newton polytopes of these $15$ polynomials, remarkably we find a polytope with $16$ facets whose $f$-vector is
\[{\bf f}=(1,49,99,66,16,1)
\]
which is almost a $D_4$ polytope (which has ${\bf f}=(1,50,100,66,16,1)$). Moreover, the (outward) normal vectors of all these $16$ facets are nothing but the $g$-vectors of the $16$ letters, which allow us to identify each letter with a facet of the polytope. Note that both co-dimension $1$ and $2$ boundaries of this polytope agree with those of $D_4$ polytope, but it misses one edge and one vertex (and becomes slightly non-simple). We can of course include the last $F$-polynomial for Minkowski sum/tropicalization, which will then give exactly the $D_4$ polytope. 

Next we consider two-mass-easy case with dual points
$(x_1, x_2, x_4, x_5, x_7, x_8)$: the (co-dimension 4) positroid is given by the two conditions above and $\langle 34 56\rangle=\langle 56 78\rangle=0$, and we have decorated permutation $\sigma=\{7, 5, 6, 9, 8, 10, 12, 11\}$ and plabic graph
\begin{center}
\includegraphics[scale=0.2]{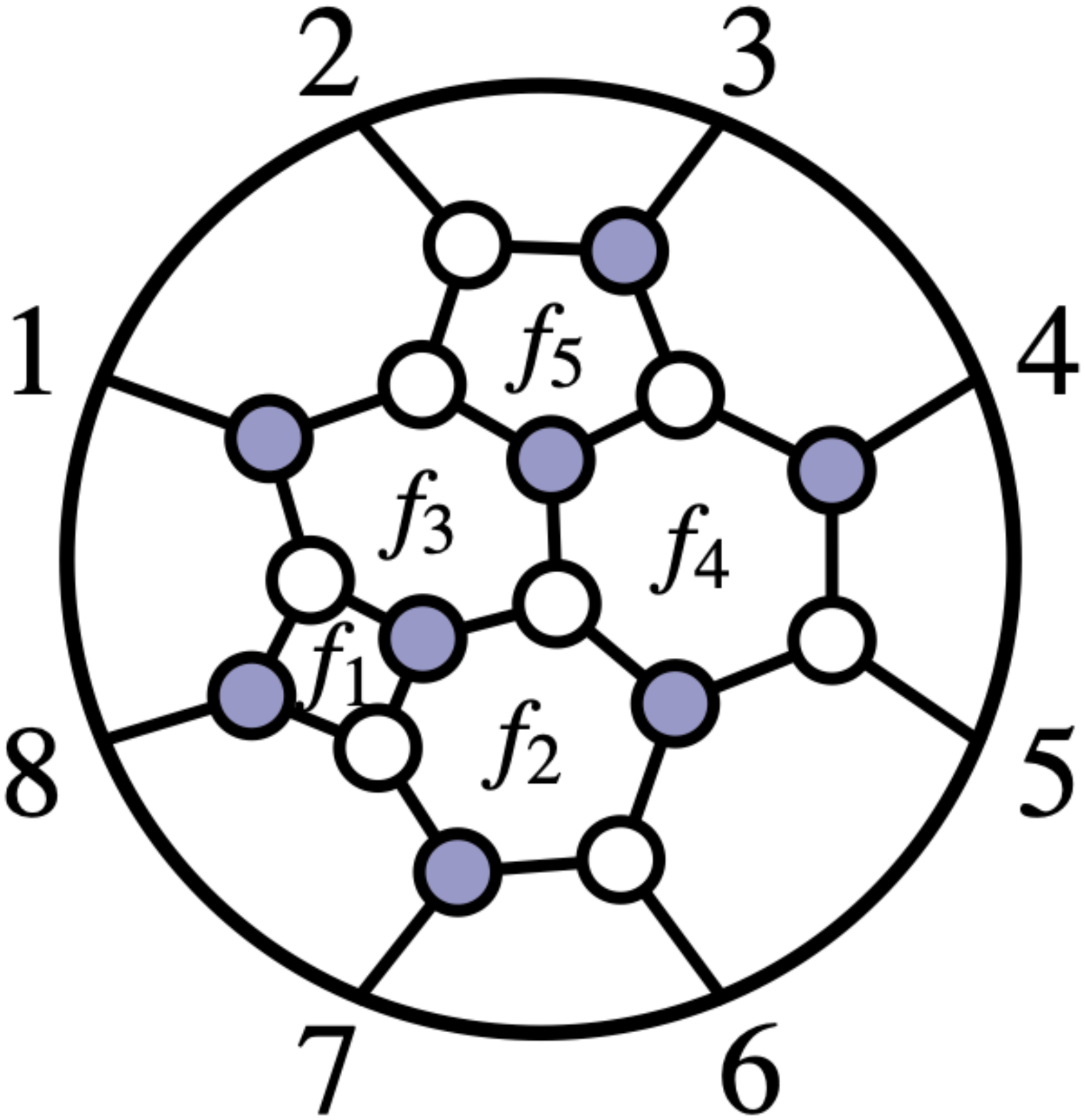}
\quad \quad 
\raisebox{.4\height}{\includegraphics[scale=0.45]{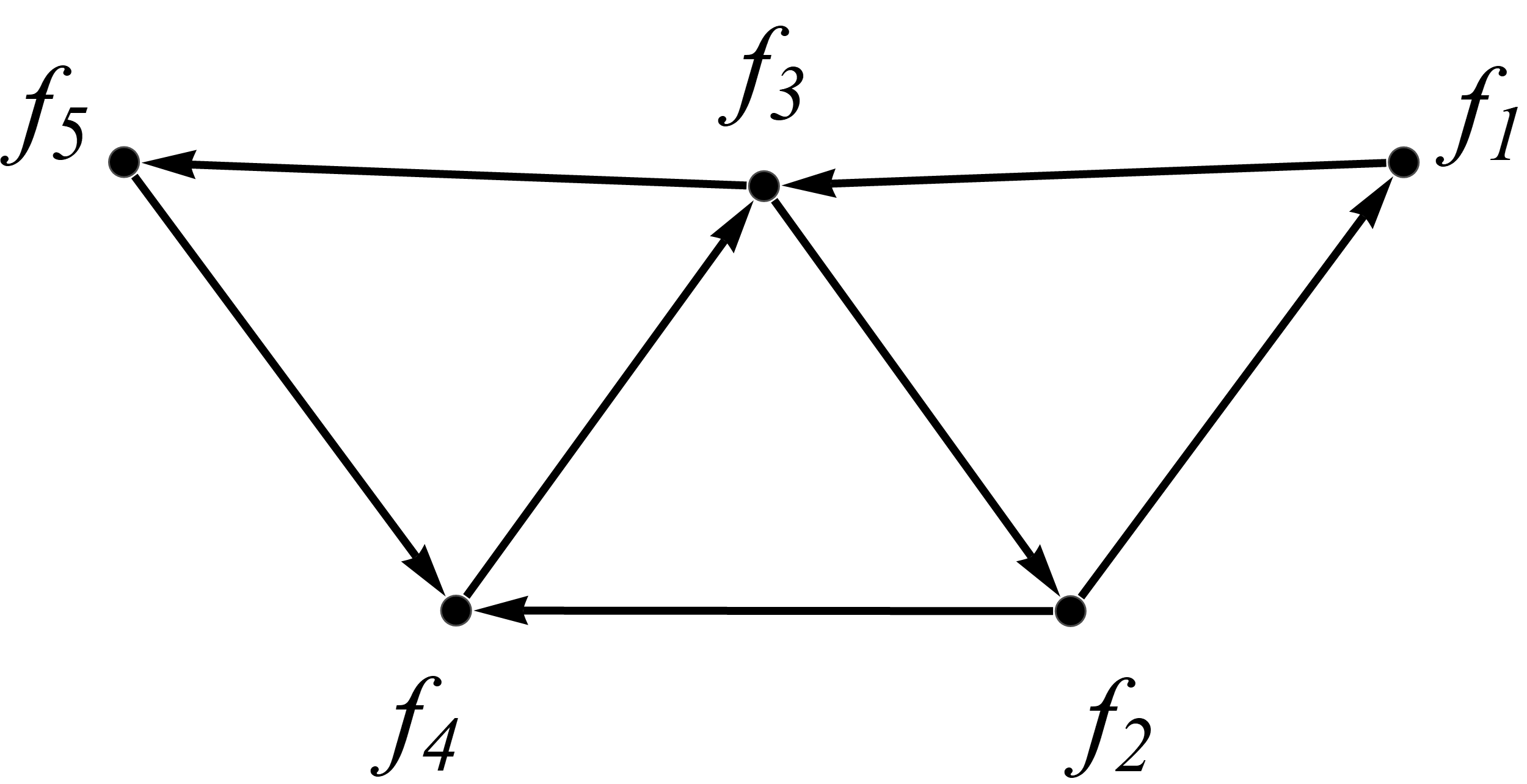}}
\end{center}
We obtain the following ${\bf Z}$ matrix as a positive parametrization (after modding out torus action):
{\footnotesize $$\left(
\begin{array}{cccccccc}
 f_5 & f_5 & 1{+}f_5 & 1 & 0 & 0 & 0 & 0 \\
 0 & f_1 f_2 f_3 f_4 f_5 & f_2 f_4 \left(1{+}f_1{+}f_1 f_3{+}f_1 f_3 f_5\right) & 1{+}f_2{+}f_1 f_2{+}f_2 f_4{+}f_1 f_2 f_4{+}f_1 f_2 f_3 f_4 & 1{+}f_2{+}f_1 f_2 & 1 & 0 & 0 \\
 0 & 0 & f_2 f_4 & 1{+}f_2{+}f_2 f_4 & 1{+}f_2 & 1 & 0 & -1 \\
 0 & 0 & 0 & 1 & 1 & 1 & 1 & 0 \\
\end{array}
\right).$$}

The dual quiver diagram (on the right) is also one for the $D_5$ cluster algebra; by applying mutation rules we find $25$ letters, including $f_1, \cdots, f_5$ and $20$ $F$-polynomials of $f$'s. We will not explicitly write this $D_5$ alphabet: it suffices to say that it consists of $23$ positive polynomials from all non-vanishing Pl\"ucker coordinates of the ${\bf Z}$ matrix above, and two missing letters, which are $1+f_3$ and $1+f_2+f_5+f_2 f_5+f_2 f_4 f_5$. 

To obtain a truncated cluster algebra, we take the Minkowski sum of Newton polytopes of the $23$ polynomials, and we obtain a polytope with $25$ facets whose $f$-vector is 
\[
{\bf f}=(1,178,449,408,160,25,1),
\]
which is a truncated $D_5$ polytope. The normal vectors of these facets turn out to be exactly the $g$-vectors of the $25$ letters we find. We see that again it differs from $D_5$ polytope starting from co-dimension $3$ boundaries, and by including the two missing $F$-polynomials we of course recover the $D_5$ polytope. 

As our last warm-up example, we consider three-mass-easy kinematics with dual points $(x_1, x_2, x_4, x_5, x_7, x_8)$: the (co-dimension 6) positroid is given by the four conditions above and additionally $\langle 67 89\rangle=\langle 89 12\rangle=0$, thus we have the decorated permutation $\sigma=\{7, 5, 6, 10, 8, 9, 13, 11, 12\}$, and the plabic graph 
\begin{center}
\includegraphics[scale=0.2]{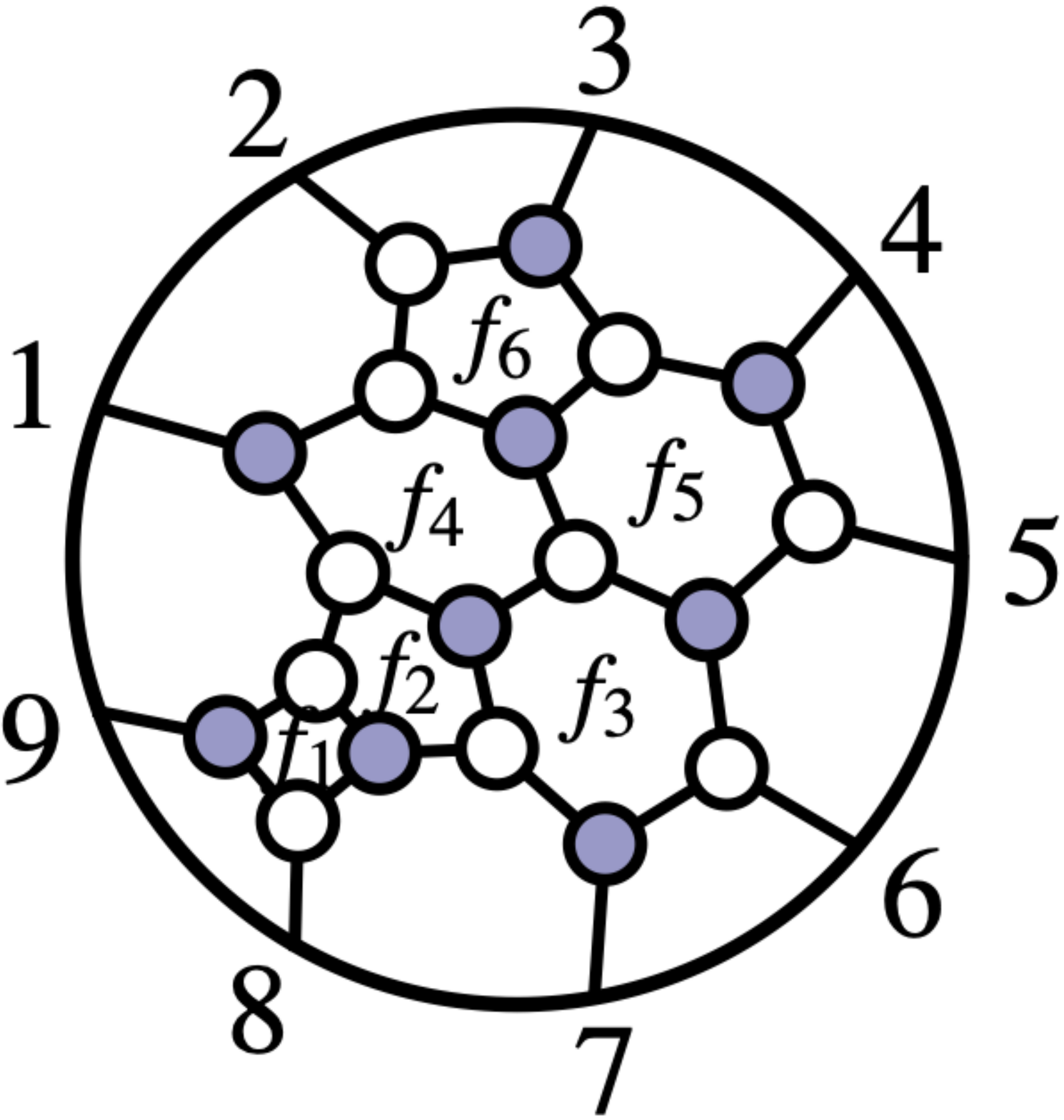}
\quad \quad 
\raisebox{.4\height}{\includegraphics[scale=0.5]{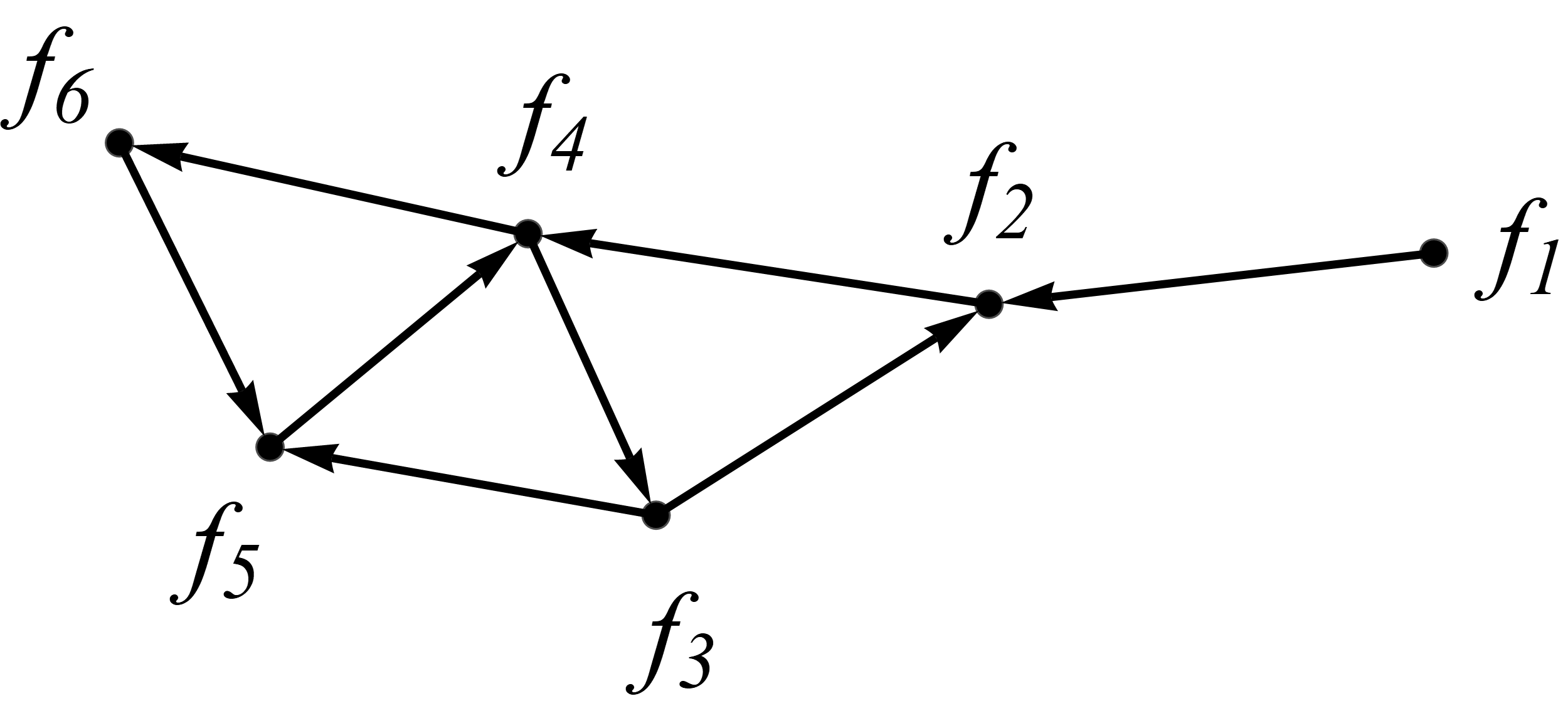}}
\end{center}
and after modding out torus action, the ${\bf Z}$ matrix reads
{\footnotesize$$\left(
\begin{array}{ccccccccc}
 f_6 & f_6 & 1+f_6 & 1 & 0 & 0 & 0 & 0 & 0 \\
 0 & f_1 f_2 f_3 f_4 f_5 f_6 & f_3 f_5 \left(1 {+}f_1 {+}f_1 f_2{+}f_1 f_2 f_4{+}f_1 f_2 f_4 f_6\right) & * & 1{+}f_1{+}f_3{+}f_1 f_3{+}f_1 f_2 f_3 & 1{+}f_1 & 0 & -1 & 0 \\
 0 & 0 & f_3 f_5 & 1+ f_3+f_3 f_5 & 1+f_3 & 1 & 0 & -1 & -1 \\
 0 & 0 & 0 & 1 & 1 & 1 & 1 & 0 & 0 \\
\end{array}
\right)$$} where $*=1{+}f_1{+}f_3{+}f_1 f_3{+}f_3 f_5{+}f_1 f_2 f_3{+}f_1 f_3 f_5{+}f_1 f_2 f_3 f_5 {+}f_1 f_2 f_3 f_4 f_5$. As we draw on the right, the quiver diagram is one for the $D_6$ cluster algebra, and the resulting alphabet consists of $f_1, \cdots, f_6$ and $30$ $F$-polynomials. They can be identified as the $33$ positive polynomials from non-vanishing Pl\"ucker coordinates of the above ${\bf Z}$ matrix, except for the following $3$:  $1+f_4$,$1+f_3+f_6+f_3 f_6+f_3 f_5 f_6 $, and $f_1 f_3 f_2^2+f_1 f_3 f_5 f_2^2+f_1 f_3 f_4 f_5 f_2^2+f_1 f_2+2 f_1 f_3 f_2+f_3 f_2+2 f_1 f_3 f_5 f_2+f_3 f_5 f_2+f_1 f_3 f_4 f_5 f_2+f_1+f_1 f_3+f_3+f_1 f_3 f_5+f_3 f_5+1$. By taking the Minkowski sum of Newton polytopes of these $33$ polynomials, we find a  truncated $D_6$ polytope with $36$ facets and $f$-vector 
\[
{\bf f}=(1,657,1986,2292,1257,330,36,1).
\]
The normal vectors agree with all the $g$-vectors of the $36$ letters, but it differs from the $D_6$ polytope starting at co-dimension $3$ boundaries. 

As checked to at least three loops in~\cite{He:2021esx}, these alphabets apply to all DCI integrals we computed with such kinematics, including double-penta-ladder integrals for $n=6,7,8$ with various possible numerators. It is remarkable that their symbol alphabets seem to be determined by truncated cluster algebras naturally associated with the kinematics.

In the next subsection, we move to a more non-trivial case, where the cluster algebra from the dual quiver is an infinite type (affine $D_4$).
The Minkowski sum becomes crucial for this case since it provides a natural truncation that gives a finite (rational) alphabet, as well as limit ray(s) that gives non-rational letters. 

\subsection{Truncated affine $D_4$ cluster algebras}

The main example we are interested in is the hexagon with two massive corners on opposite sides, where we have dual points $(x_1, x_2, x_4, x_5, x_6, x_8)$. The (co-dimension $4$) positroid can be obtained by $\langle 8123\rangle=\langle 2345\rangle=\langle 4567\rangle=\langle 6781\rangle=0$, thus the decorated permuation reads $\sigma=\{6, 5, 8, 7, 10, 9, 12, 11\}$. We have a rather symmetric plabic graph,
\begin{figure}[htbp]
\centering
\includegraphics[scale=0.2]{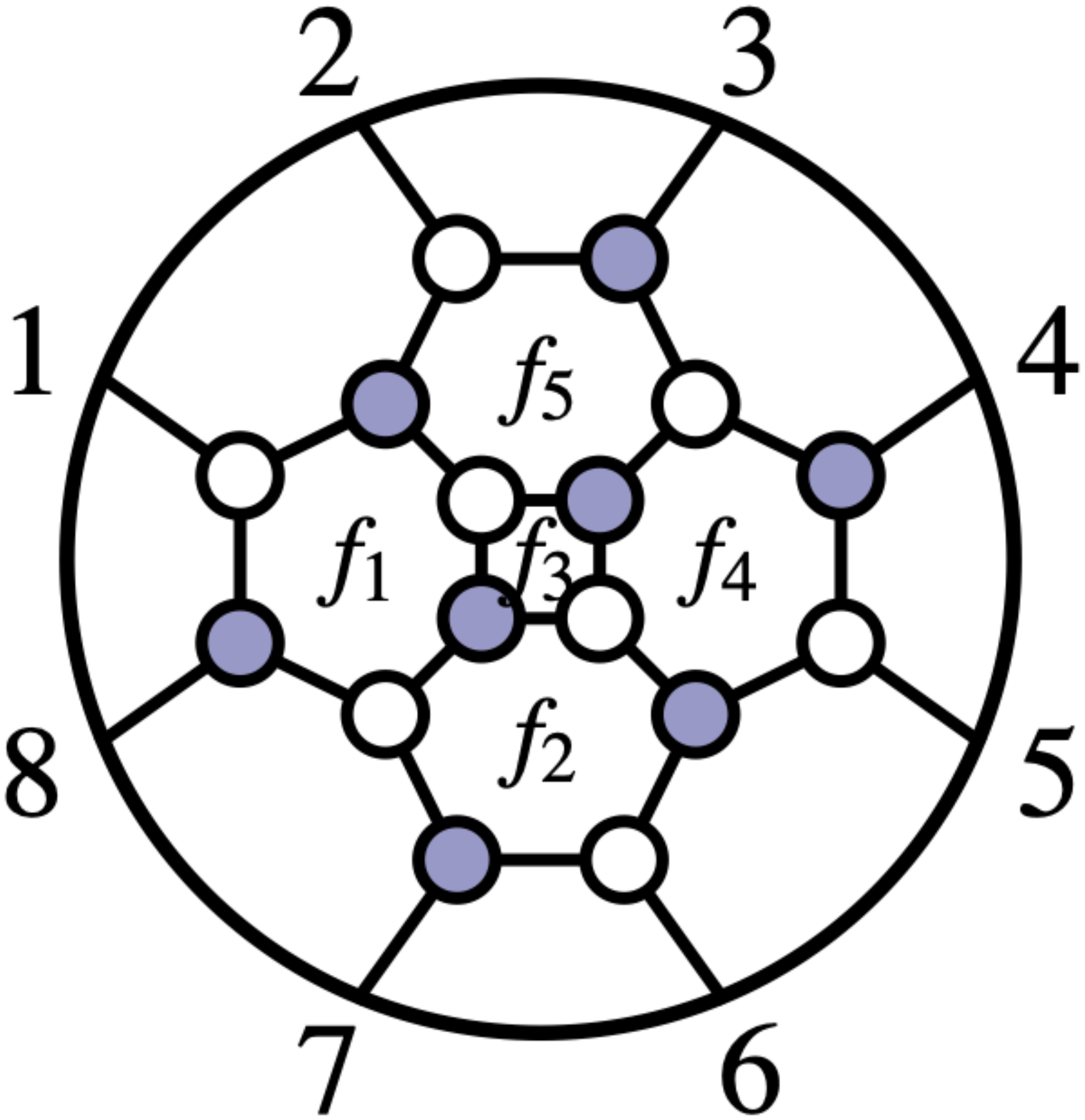}
\quad \quad 
\raisebox{.18\height}{\includegraphics[scale=0.32]{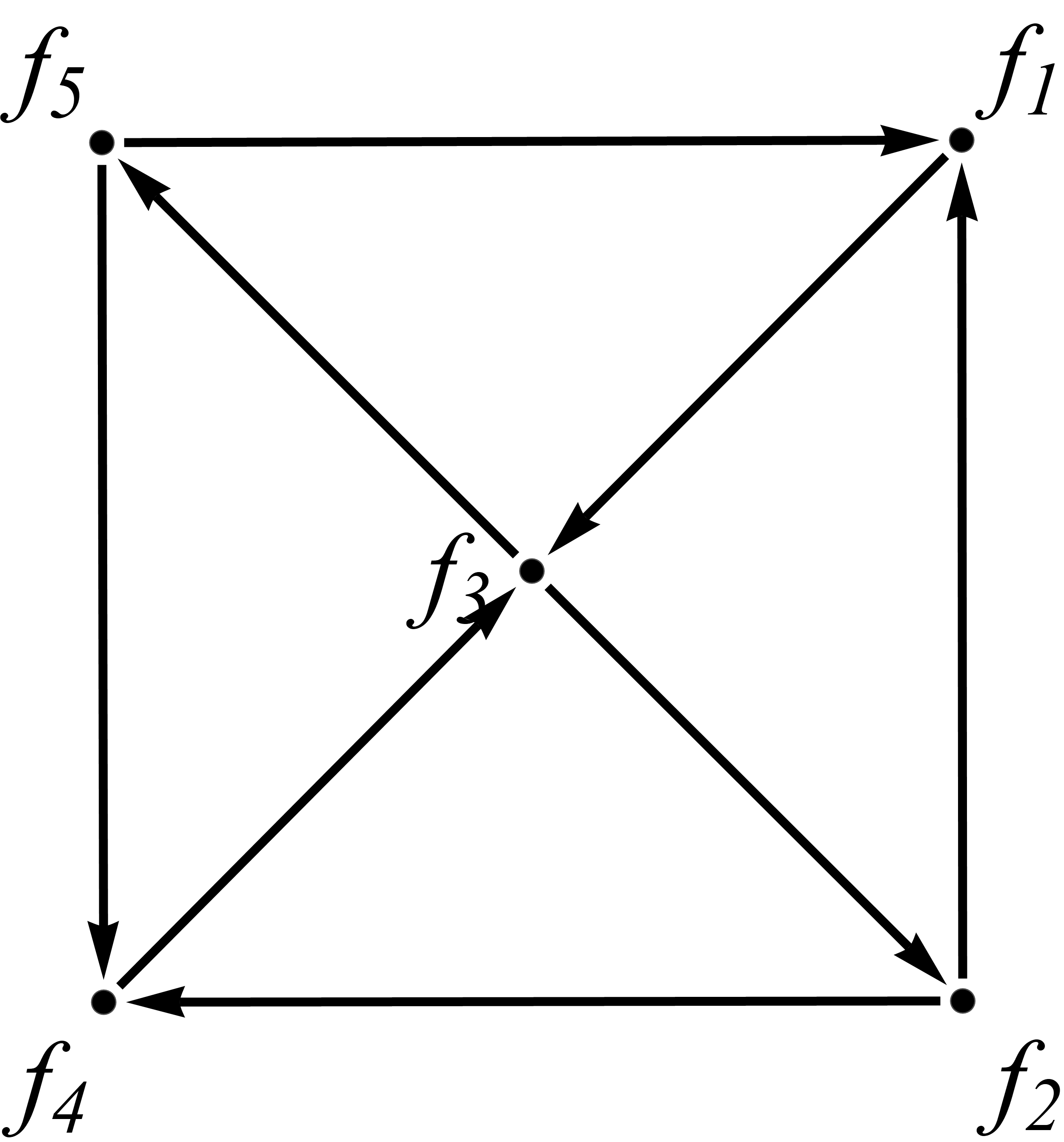}}
\end{figure}
and the dual quiver diagram can be identified with one for affine $D_4$ type. This is an infinite-type cluster algebra (though it is mutation finite), and we must rely on Minkowski sum to obtain a finite alphabet. After modding out the torus action, we have the ${\bf Z}$ matrix:
$$\left(
\begin{array}{cccccccc}
 f_1 f_2 f_3 f_4 f_5 & f_1 f_2 f_3 f_4 f_5 & f_2 f_4 \left(-1+f_1 f_3 f_5\right) & -1-f_2-f_2 f_4 & -1-f_2 & -1 & 0 & 0 \\
 0 & f_3 f_4 f_5 & f_4 \left(1+f_3 +f_3 f_5\right) & 1+f_4 +f_3 f_4 & 1 & 0 & 0 & 0 \\
 0 & 0 & f_2 f_4 & 1+f_2+ f_2 f_4 & 1+f_2 & 1 & 0 & -1 \\
 0 & 0 & 0 & 1 & 1 & 1 & 1 & 0 \\
\end{array}
\right)$$

From all non-vanishing Pl\"ucker coordinates, we find exactly $25$ positive polynomials, which we record as $W_i$ for $i=1, \cdots, 25$ (anticipating that they will be part of the full alphabet).  The first $10$ of them are linear in $f$'s: which we write as 
\begin{equation}
W_i=f_i, \quad W_{5+i}=1+f_i,\qquad {\rm for}~i=1,\cdots, 5;
\end{equation}
The next $8$ letters are degree-$2$ polynomials of the form $w_{i,j}:=1+f_j+ f_i f_j$:
\begin{align}
&W_{11}=w_{1, 2}, \,\,W_{12}=w_{3,1}, \,\,W_{13}=w_{2,3}, \,\,W_{14}=w_{4,2}, \nonumber\\
&W_{15}=w_{3,4}, \,\,W_{16}=w_{1,5}, \,\,W_{17}=w_{5,3}, \,\,W_{18}=w_{4,5}.
\end{align}
Finally, the last $7$ letters involve polynomials of degree $3, 4$ or $5$; introducing $w_{i,j,k}:=1+f_i+ f_j+f_i f_j + f_i f_j f_k$; we have 
\begin{align}
&W_{19}=w_{1,4,3}, W_{20}=1+f_3(f_2+w_{2,5}), W_{21}=1+f_2 w_{1,4,3}, W_{22}=1+f_3 w_{2,5,1}, \\
&W_{23}=1+ f_5 w_{1,4,3},\,\, W_{24}=1+ f_3 w_{2,5,4},\,\, W_{25}=1+f_3 (w_{2,5,1}+ w_{3,1} f_2 f_4 f_5).\nonumber
\end{align}

By taking the Minkowski sum of their Newton polytopes, we obtain a $5$-dimensional polytope with $f$-vector 
\[{\bf f}=(1,280, 739, 694, 272, 39,1),\]
and it is easy to compute the normal vectors of these $39$ facets.
By comparing these $39$ vectors with $g$-vectors of the affine $D_4$ cluster algebra above, we see that $38$ of them correspond to $g$-vectors, and for completeness we record them here. For $W_1, \cdots, W_5$, their $g$-vectors are $g_i={\bf e}_i$ for $i=1,\cdots, 5$, and the remaining $33$ $g_i$ for $i=6, 7, \cdots, 38$ read:

{\footnotesize \begin{align}
& (-1 , 0 , 1 , 0 , 0 ),(1 , -1 , 0 , 1 , 0 ),(0 , 1 , -1 , 0 , 1 ),(0 , 0 , 1 , -1 , 0 ),(1 , 0 , 0 , 1 , -1 ),(0 , -1 , 0 , 1 , 0 ),(-1 , 0 , 0 , 0 , 0 ),\nonumber\\
& (0 , 0 , -1 , 0 , 1 ), (1 , -1 , 0 , 0 , 0 ), (0 , 0 , 0 , -1 , 0 ),(0 , 0 , 0 , 1 , -1 ),(0 , 1 , -1 , 0 , 0 ),(1 , 0 , 0 , 0 , -1 ),(-1 , 0 , 1 , -1 , 0 ),\nonumber\\
& (1 , 0 , -1 , 1 , 0 ),(0 , -1 , 0 , 0 , 0 ),(0 , 0 , -1 , 1 , 0 ),(0 , 0 , 0 , 0 , -1 ),(1 , 0 , -1 , 0 , 0 ),(0 , 0 , -1 , 0 , 0 ),(0 , -1 , 1 , 0 , 0 ), \nonumber\\
& (1 , -1 , 0 , 2 , -1 ),(0 , 0 , 1 , 0 , -1 ),(2 , -1 , 0 , 1 , -1 ),(0 , -1 , 1 , 1 , -1 ),(0 , -1 , 0 , 1 , -1 ),(1 , -1 , -1 , 1 , 0 ),\nonumber\\
& (1 , -1 , 1 , 0 , -1 ),(1 , -1 , 0 , 0 , -1 ),(0 , -1 , 1 , 0 , -1 ),(1 , -2 , 0 , 1 , -1 ),(1 , 0 , -1 , 1 , -1 ),(1 , -1 , 0 , 1 , -2 )\nonumber
\end{align}}

These $38$ facets then give $F$-polynomials including the above $25$ polynomials, and we find additionally $13$ polynomials. We denote these letters as $W_{26}, \cdots, W_{38}$.  Note that some of the remaining ones are relabelling of what we have seen in the first $25$ letters. For example, $W_{26}=1+f_2(f_1+w_{1,4})$, $W_{27}=w_{2,5,1}$, $W_{28}=1+f_5(f_1+w_{1,5})$, {\it etc.}. All $38$ rational letters are recorded in the ancillary file. Note that these letters can also be obtained by simply parametrizing the $356$ rational letters of $G_+(4,8)/T$ using our $\bf{Z}$ matrix. It is interesting to see that if we start with the  smaller (rational) alphabet with $272$ letters for $G_+(4,8)/T$, we obtain only $33$ letters with $\{W_{30},W_{33},W_{35},W_{36},W_{38}\}$ missing, and we will come back to this smaller alphabet later.  It is, however, not clear to us how to directly obtain the $33$ letters (plus algebraic ones) by Minkowski sum; {\it e.g.} if we use parity-invariant subset of non-vanishing minors of our $\bf{Z}$ matrix, we obtain a polytope with only $18$ facets and all of them correspond to rational letters, which is insufficient.

There is a remaining normal vector, $g_\infty=(1, -1, 0, 1, -1)$.  After extensive search, it turns out not to be any $g$-vector of the infinite cluster algebra. As shown in \cite{Drummond:2019cxm},  after infinite sequences of mutations on a quiver with doubled arrow 
, the directions of $g$-vectors on two ends of the doubled arrow will asymptote to the so-called {\it limit ray}. Difference between the two g-vectors on the end of doubled arrow will stay invariant in the infinite mutations, giving the limit rays they asymptote to, which is exactly our $g_\infty$! For instance, after mutation series $\{5,1,4\}$ from the initial cluster, the quiver turns out to be
\begin{center}
{\includegraphics[scale=0.45]{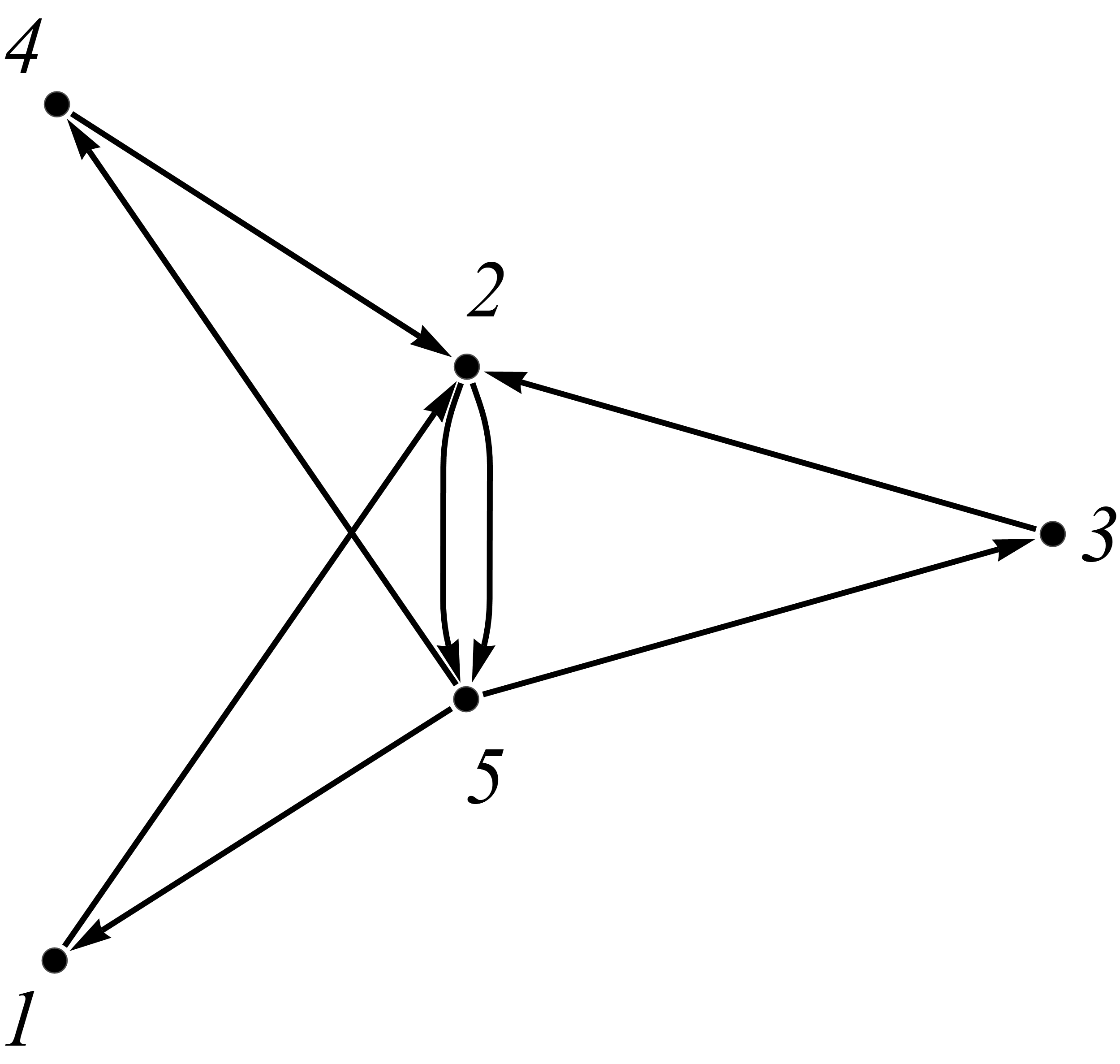}},
\end{center}
with $g(2)=g_2$, $g(5)=g_{10}$. It is straightforward to check that $g_\infty=g_{10}-g_{2}$, and the difference stays invariant in the infinite mutations.

As can be computed from the algorithm in \cite{Drummond:2019cxm, Henke:2019hve, Arkani-Hamed:2019rds}, $g_\infty$ is associated with exactly the square root for the unique four-mass-box $(x_2, x_4, x_6, x_8)$ for this kinematics, which is defined as
\begin{equation}
\Delta:=\sqrt{(1-u_3-v_3)^2-4 u_3 v_3}, \quad u_3{=}\frac{\langle1234\rangle\langle5678\rangle}{\langle1256\rangle\langle3478\rangle},\ v_3{=}\frac{\langle1278\rangle\langle3456\rangle}{\langle1256\rangle\langle3478\rangle},
\end{equation}
where the two cross-ratios can be expressed using the letters as $u_3= 1/W_{25}, v_3=W_1 W_2 W_3^2 W_4 W_5/W_{25}$.  It is convenient to introduce the two roots $\alpha_\pm =\frac 1 2 (1+u_3-v_3 \pm \Delta)$ (such that $\alpha_+-\alpha_-=\Delta$), which appear in the (second entry of) symbol of the famous four-mass box 
$${\cal S}[F(x_2, x_4, x_6, x_8)]=-\frac 1 2 \left(v_3 \otimes L_1 + u_3 \otimes L_2\right),$$
where the two simplest {\it algebraic letters} are denoted as $L_1=\frac{\alpha_+}{\alpha_-}$ and $L_2=\frac{1-\alpha_-}{1-\alpha_+}$. In addition, we find infinite sequences of mutations which produce these and other algebraic letters, similar to what was done in~\cite{Drummond:2019cxm, Henke:2019hve}. The upshot is that we find a space of $5$ multiplicative independent algebraic letters: $L_1, L_2$ and 
\begin{equation}
L_3=\frac{W_{17}^{-1}-\alpha_-}{W_{17}^{-1}-\alpha_+}, \quad L_4=\frac{W_{13}/W_{25}-\alpha_-}{W_{13}/W_{25}-\alpha_+},\quad L_5=\frac{(1-W_1 W_2 W_3)^{-1}-\alpha_+}{(1-W_1 W_2 W_3)^{-1}-\alpha_-}\,.
 \end{equation}
It is remarkable that this is precisely the $5$-dimensional space of algebraic letters found for double-pentagon integral $\Omega_2(1,4,5,8)$~\cite{He:2020lcu}!

\section{The cluster function space and double-penta ladders to four loops}\label{sec:3}

\subsection{First entries, cluster adjacency and algebraic letters}\label{sec:3.1}

Having obtained the alphabet with $38$ rational letters and $5$ non-rational ones, it is natural to construct the space of cluster functions, and we will content ourselves with first building all integrable symbols. There are two important constraints we can impose: first, we are interested in symbols whose first entries consist of only physical discontinuities, which can be chosen to be $5$ independent space-time cross-ratios. Moreover, we will impose cluster adjacency conditions, {\it i.e.} only letters that appear in the same cluster (of the truncated cluster algebras) can be adjacent to each other in the symbol. 

As discussed in~\cite{He:2021esx}, the $5$ independent cross-ratios which can appear on the first entry are $u_3, v_3$ defined above, as well as the following three: 
\[u_1{=}\frac{\langle1245\rangle\langle5681\rangle}{\langle1256\rangle\langle4581\rangle}=\frac{1}{W_8}, \ u_2{=}\frac{\langle3481\rangle\langle4578\rangle}{\langle3478\rangle\langle4581\rangle}=\frac{W_{13} W_{17}}{W_8 W_{25}},\ u_4{=}\frac{\langle1234\rangle\langle4581\rangle}{\langle1245\rangle\langle3481\rangle}=\frac{W_8}{W_{17}}\]

With the alphabet and first entries, we are ready to build functions or integrable symbols, starting from the $\log(u_1), \log (u_2), \log(u_3), \log(v_3), \log(u_4)$ at weight $1$. Our construction is recursive: at each weight $w$, we consider all integrable symbols of weight $w-1$ tensored with any of the $38+5$ letters, and impose integrability conditions on the final two entries. We start from the ansatz $\sum_{i,j} c_{i,j} S^{(w{-}1)}_i d\log l_j$ where $S^{(w{-}1)}_i$ denotes weight-$(w{-}1)$ integrable symbols, and $l_j$ the letters, {\it i.e.} $W_1, \cdots, W_{38}$, $L_1, \cdots, L_5$. The integrability condition reads
\begin{equation}\label{integrability}
\sum_{i, j, m} c_{i,j} S^{(w{-}2)}_{i;m} d\log l_m \wedge d\log l_j=0,
\end{equation}
where $S^{(w{-}2)}_{i;m}$ denote coefficients of $d\log l_m$ in $S^{(w{-}1)}_i$, which are linear combinations of weight-$(w{-}2)$ integrable symbols. Therefore, all we need is to find all linear relations among ${43 \choose 2}$ $d\log$ $2$-forms (some of them vanish identically, {\it e.g.} $d\log f_i \wedge d\log (1+f_i)=0$), and all such relations are recorded in the ancillary file. In this way, we can easily construct the space to relatively high weight: it turns out that there are $5, 24, 113, 530$ such integrable symbols at weight $w=1,2,3,4$. 

Now we turn to possible cluster adjacency conditions to reduce the space, which forbid letters that cannot be in the same cluster to appear next to each other in the symbol. More precisely, we will use the truncated cluster algebra and its polytope for imposing these conditions: if two letters have facets that intersect in the polytope, then clearly they belong to the same cluster, otherwise we claim that they are a {\it forbidden pair} in the truncated cluster algebra. We do not know if there exists a cluster in the infinite affine $D_4$ cluster algebra which includes a forbidden pair, but for our purpose we will use this ``truncated" version of cluster adjacency and forbid such a pair to appear next to each other in the symbol~\footnote{This may sounds too strong, but in fact what we have done is we first ``bootstrapped" the integrals $\Omega_L(1,4,5,8)$ up to $L=4$ {\it without} using such adjacency conditions; the result does respect these conditions, which means that they can indeed be imposed, and in the following we present the improved bootstrap in this reduced space.}. 

We apply this version of cluster adjacency to the rational letters $W_1, \cdots, W_{38}$ (it is not clear to us  how to extend it to include the remaining $5$ $L$'s which are assigned to the same facet). Since $W_1, \cdots, W_5$ are not $F$-polynomials, we do not consider them in the study of adjacency conditions; in other words, we list the facets for $W_i$ with $i=6,7, \cdots, 38$, and find all pairs that do not intersect in the polytopes. In this way, we find $350$ forbidden pairs out of $\frac{33\times 34}2$, which we record in the ancillary file (in practice, this can be trivially done by using {\it e.g.} {\bf polymake}). By applying these adjacency conditions to the construction, and we find that the space is reduced significantly (more and more so for higher weights).  For $w=1,2,3,4$, the dimension of the space is reduced to $5, 23, 93, 340$. Moreover, we have computed the reduced space for $w=5,6$, and find $1141, 3585$ such integrable symbols respectively. The physical meaning of such adjacency is unclear as usual, but we conjecture that this reduced space contains DCI Feynman integrals with such ``two-mass opposite" kinematics, and we can use it to bootstrap such integrals at least up to three loops. 

Before proceeding, we remark that similarly one can bootstrap for these warm-up cases such as $D_4$ functions for one-mass hexagon kinematics with $n=7$. Note that this $D_4$ alphabet can be obtained as a boundary of our truncated cluster algebra {\it e.g.} by sending $u_3 \to 0$. It is straightforward to construct the space of integrable symbols for $D_4$ functions, with first entries given by $u_1, u_2, u_4, v_3$. The dimensions of the space at weight $1,2,3,4$ are $4, 16, 63, 246$; we can impose adjacency conditions which forbid $30$ pairs out of $\frac{12 \times 13} 2$ pairs of $F$-polynomials, and these conditions reduce the dimensions to $4, 15, 50, 155$ up to weight $4$. Nicely, any integral with one-mass hexagon kinematics up to weight $4$ that we know of can be found in the space. 

What can we say about non-rational letters? Although we do not know how to impose conditions such as cluster adjacency on them, it turns out that they are still constrained at least at low weights. The first observation is that there is only one weight $2$ function involving them: the four-mass box whose symbol we record above, where we have $L_1$ and $L_2$ in the second entry. Similarly we find only $11$ weight $3$ functions with algebraic letters. Among them, the first five have the  form
\begin{equation}\label{even}
    \mathcal{S}(F(x_2,x_4,x_6,x_8))\otimes L_i+{\rm rational\ part}
\end{equation}
with $i=1\cdots 5$. While other $6$ functions are linear combinations of
\begin{equation}\label{odd}
    \{\mathcal{S}(F(x_2,x_4,x_6,x_8))\otimes W_j, \quad \mathcal{S}({\rm dilog.\ with}~W_j)\otimes L_i\}
\end{equation}
Note that under a ``parity" $\Delta \to -\Delta$, those symbols in \eqref{even} stay invariant, while those in \eqref{odd} picks up a minus sign. For any ``parity-even" amplitude or integral, what we need are those even functions in \eqref{even}, or those odd ones in \eqref{odd} dressed with a prefactor that is a odd function in $\Delta$, such as $1/\Delta$~\footnote{This is also true for the ``odd" four-mass box at weight $2$, which can be normalize with a prefactor $1/\Delta$ to make it ``even", {\it e.g.} when appearing in one-loop amplitudes.}.

For higher weights, the number of functions involving algebraic letters grows rapidly. However, we are mostly interested in a particular class of functions starting at weight $4$. In the next subsection, we will locate $\Omega_L(1,4,5,8)$ up to $L=4$, and for now let's see how much it takes to determine the part that contains algebraic letters at $L=2$. We will show shortly how to determine the {\it last entries} of  $\Omega_L(1,4,5,8)$ for $L\geq 2$ from the Wilson-loop $d\log$ form or differential equations: starting $L=2$, the symbol of $\Omega_L(1,4,5,8)$ contains exactly $5$ last entries, which we denote as $\{z_i\}_{i=1\cdots 5}$: 
\begin{equation}\label{lastentry}
z_1=-W_3, z_2=-\frac{W_2 W_3 W_5 W_{12} W_{15}}{W_{13} W_{17}}, z_3=\frac{W_2 W_3}{W_{13}}, z_4=\frac{W_3 W_5}{W_{17}}, z_5=\frac{W_2 W_3^2 W_5 W_{12}}{W_{13} W_{17}}
\end{equation}
Therefore it is natural to see what symbols with algebraic letters and only these $5$ last entries can we find in the space. Surprisingly after imposing last-entry conditions on weight $4$,  only {\it one} independent weight-$4$ functions containing algebraic letters $L_i$ is left, and we record this {\it integrable} symbol up to the part involving purely rational letters $W_j$'s (the rational part depends on our basis of weight-$4$ functions): 
{\small\begin{equation}
{\cal S}_{2,4,6,8}:=\mathcal{S}(F_{2,4,6,8})\otimes(\frac{L_2L_5}{L_1L_3}\otimes z_1+\frac{L_2L_5}{L_1L_4}\otimes z_2+\frac{L_5}{L_1^2L_3L_4}\otimes z_3+\frac{L_5}{L_1}\otimes z_4+\frac{L_1^2L_3L_4}{L_2L_5^2}\otimes z_5)+ {\rm rational}\label{algpart}
\end{equation}}
where we have denoted the four-mass box as $F_{2,4,6,8}:=F(x_2, x_4, x_6, x_8)$. We see that by restricting to the five last entries, exactly the first five weight-$3$ functions described above contribute, which can be viewed as generating the first derivatives $\partial_{z_i}$ for $i=1,2,\cdots, 5$ of ${\cal S}_{2,4,6,8}$. In fact, the first two weight-$3$ functions (involving $L_1$ and $L_2$) can be chosen to be the two weight-$3$ functions that appear when solving differential equation for double-box integral~\cite{Drummond:2010cz}, and it is nice to see that we just have three additional weight-$3$ functions involving $L_3, L_4, L_5$, when solving weight-$4$ double-pentagon. We record the symbol of $\Omega_2(1,4,5,8)$ in ancillary file, and it is easy to see that the algebraic part is given by \eqref{algpart}.

Having obtained a function that captures the algebraic part of $\Omega_2(1,4,5,8)$, we remark that from it we can easily obtain the algebraic part of the most general double-pentagon integral; we denote it as $\Omega_2(i,j,k,l)$ with the first fully general case for $n=12$ and {\it e.g.} $(i,j,k,l)=(1,4,7,10)$~\cite{He:2020lcu}.
$\Omega_2(i,j,k,l)$ contains $2^4=16$ four-mass-box square roots, labelled by $(x_a,x_b,x_c,x_d)$ with $(a,b,c,d)=(i+\sigma_i, j+\sigma_j, k+\sigma_k, l+\sigma_l)$ with $\sigma=0,1$~\cite{He:2020lcu}. For each $(a,b,c,d)$, all we need to do is simply relabel the momentum twistors of $\Omega_2 (1,4,5,8)$ by $\{1\to i,4\to j,5\to k,8\to l\}$ and $\{2\to i\pm1,3\to j\pm1,6\to k\pm1,7\to l\pm1\}$ where the choice $\pm 1$ depends on $\sigma$'s, {\it e.g.} for $\sigma_i=1$ ($a=i+1$), $2\to i+1$. By summing over $16$ such relabelled symbol (with alternating signs), we obtain an integrable symbol that contains the algebraic part of $\Omega_2(i,j,k,l)$:
\begin{equation}
    {\cal S}(\Omega_2(i,j,k,l))=\sum_{\{\sigma\}} (-)^{\sum \sigma} {\cal S}_{i+\sigma_i, j+\sigma_j, k+\sigma_k, l+\sigma_l} + {\cal S}(R)
\end{equation}
where the sum is over $2^4=16$ choices of $\sigma$'s with a minus sign when  $\sigma_i + \sigma_j + \sigma_k + \sigma_l$ is odd; $R$ denotes a weight-$4$ function with only rational letters. It is remarkable that, up to this $R$ function, the most generic double-pentagon integral can be obtained using $16$ weight-$4$ integrable symbols found in our space.

\subsection{Double-penta-ladders: last entries, differential equations {\it etc.}}\label{sec:3.2}

Now we move to the computation of double-penta ladder integrals, $\Omega_L(1,4,5,8)$, which can be defined directly from Wilson loop $d\log$ representation: we can rewrite an $L$-loop ladder as a two-fold integral over a $(L-1)$-loop integral as
\begin{equation}
  \Omega_L(1,4,5,8)=\int d\log{\langle 148Y\rangle}\,  d\log\frac{\langle1X4Y\rangle}{t}\times\begin{tikzpicture}[baseline={([yshift=-.5ex]current bounding box.center)},scale=0.18]
        \draw[black,thick] (0,0)--(0,5)--(4.76,6.55)--(7.69,2.5)--(4.76,-1.55)--cycle;
        \draw[black,thick] (-15,5)--(-19.76,6.55)--(-22.69,2.5)--(-19.76,-1.55)--(-15,0);
        \draw[decorate, decoration=snake, segment length=12pt, segment amplitude=2pt, black,thick] (4.76,6.55)--(4.76,-1.55);
        \draw[decorate, decoration=snake, segment length=12pt, segment amplitude=2pt, black,thick] (-19.76,6.55)--(-19.76,-1.55);
        \draw[black,thick] (9.43,1.5)--(7.69,2.5)--(9.43,3.5);
        \draw[black,thick] (4.76,6.55)--(5.37,8.45);
        \draw[black,thick] (4.76,-1.55)--(5.37,-3.45);
        \draw[black,thick] (0,5)--(-5,5)--(-5,0)--(0,0);
        \draw[black,thick,densely dashed] (-5,5)--(-10,5);
        \draw[black,thick,densely dashed] (-5,0)--(-10,0);
        \draw[black,thick] (-10,0)--(-10,5)--(-15,5)--(-15,0)--cycle;
        \draw[black,thick] (-19.76,6.55)--(-20.37,8.45);
        \draw[black,thick] (-19.76,-1.55)--(-20.37,-3.45);
        \draw[black,thick] (-24.69,3.5)--(-22.69,2.5)--(-24.69,1.5);
        \filldraw[black] (-20.37,8.45) node[anchor=south] {{4}};
        \filldraw[black] (5.37,8.45) node[anchor=south] {{5}};
        \filldraw[black] (9.43,3.5) node[anchor=west] {{6}};
        \filldraw[black] (9.43,1.5) node[anchor=west] {{7}};
        \filldraw[black] (5.37,-3.45) node[anchor=north] {{8}};
        \filldraw[black] (-20.37,-3.45) node[anchor=north] {{1}};
        \filldraw[black] (-24.69,1.5) node[anchor=east] {{X}};
        \filldraw[black] (-24.69,3.5) node[anchor=east] {{Y}};
    \end{tikzpicture}
\end{equation}
where $X=Z_1- t Z_3$, $Y=Z_3- sZ_5$ with $t$ and $s$ integrated on $\mathbb{R}^2_{\geq0}$. We can rescale $t$ and $s$ to make DCI property manifest, and we arrive at the recursion
\begin{equation}\label{udeformation}
{\small
    \begin{split}
        \Omega_{L+\frac12}(u_1,u_2,u_4,u_3,v_3)&=\int_0^\infty d\log\frac{t{+}1}{t}\,\Omega_{L}\biggl(\frac{u_1(t{+}u_4)}{t{+}u_1 u_4},u_2,\frac{u_4(t{+}1)}{t{+}u_4},\frac{u_3(t{+}1)}{t{+}u_1 u_4},\frac{t v_3}{t{+}u_1 u_4}\biggr),\\
        \Omega_{L{+}1}(u_1,u_2,u_4,u_3,v_3)&=\int_0^\infty  d\log(s{+}1)\,\Omega_{L+\frac12}\biggl(u_1,\frac{u_2(s{+}1)}{u_2s{+}1},\frac{s{+}u_4}{s{+}1},\frac{u_3(1{+}s/ u_4)}{1{+}s u_2},\frac{v_3}{1{+}s u_2}\biggr),
    \end{split}}
\end{equation}
Note that at the limit $u_3\to0$, $\Omega_L(1,4,5,8)$ and the recursions degenerate to the $\Omega_L(1,4,5,7)$ case. The source of the recursion is the one-loop $8$-pt chiral hexagon whose result is well known \cite{He:2020uxy} ({\it e.g.} in box expansion including $F_{2,4,6,8}$):
\begin{equation}\label{double}
\Omega_1(1,4,5,8)=\begin{tikzpicture}[baseline={([yshift=-.5ex]current bounding box.center)},scale=0.18]
            \draw[black,thick] (0,0)--(4,0)--(6,3.46)--(4,6.93)--(0,6.93)--(-2,3.46)--cycle;
            \draw[black,thick] (0,6.93)--(-1,8.66);
            \draw[black,thick] (4,6.93)--(5,8.66);
            \draw[black,thick] (7.74,2.46)--(6,3.46)--(7.74,4.46);
            \draw[black,thick] (4,0)--(5,-1.73);
            \draw[black,thick] (0,0)--(-1,-1.73);
            \draw[black,thick] (-4,2.46)--(-2,3.46)--(-4,4.46);
            \draw[decorate, decoration=snake, segment length=12pt, segment amplitude=2pt, black,thick] (0,7)--(0,0);
            \draw[decorate, decoration=snake, segment length=12pt, segment amplitude=2pt, black,thick] (4,0)--(4,6.93);
            \filldraw[black] (-1,8.66) node[anchor=south east] {{4}};
            \filldraw[black] (5,8.66) node[anchor=south west] {{5}};
            \filldraw[black] (7.74,4.46) node[anchor=west] {{$6$}};
            \filldraw[black] (7.74,2.46) node[anchor=west] {{$7$}};
            \filldraw[black] (5,-1.73) node[anchor=north west] {{8}};
            \filldraw[black] (0,-1.73) node[anchor=north east] {{1}};
            \filldraw[black] (-4,4.46) node[anchor=east] {{$3$}};
            \filldraw[black] (-4,2.46) node[anchor=east] {{$2$}};
        \end{tikzpicture},
\end{equation}

Note that at two loops, after some tedious calculation based on rationalization, the symbol (and even function~\cite{Bourjaily:2018aeq}) of  $\mathcal{S}(\Omega_2)$ can be computed from the recursion \eqref{double}. Its alphabet consists of $5$ algebraic letters $\{L_1,L_2,L_3,L_4,L_5\}$ and $21$ rational letters which are $\{W_1,\dots,W_{25}\}$ with $\{W_6,W_9,W_{22},W_{24}\}$ absent. As mentioned, the last entries of the answer are the five $z$-variables \eqref{lastentry}, which are related to the cross ratios $\{u_1,u_2,u_3,u_4,v_3\}$ by
\[
    u_1=\frac1{1{-}z_1},\ u_2=\frac1{1{-}z_2},\ u_4=1{-}z_4,\ 
    u_3=\frac{(1{-}z_3)(1{-}z_4)}{(1{-}z_1)(1{-}z_2)},\ v_3=-\frac{(z_1 z_2{-}z_5)(z_3 z_4{-}z_5)}{(1{-}z_1)(1{-}z_2)z_5}.
\]
These $z$-variables make many properties of the ladder integrals $\Omega_L(1,4,5,8)$ manifest, and we will use them extensively in the following discussions. For instance, the integrals have two axial symmetries, which are given by
\[
z_1 \leftrightarrow z_2 \quad \text{and}\quad  z_3\leftrightarrow z_4.
\]
The deformations are also simplified in terms of $z$-variables to
\begin{align}\label{zdeformation1}
    \Omega_{L+\frac12}(z_1,\dots,z_5)=\int_0^\infty d\log\frac{t+1}{t}\,\,\Omega_{L}\biggl(\frac{t z_1}{t-z_4+1},z_2, z_3,\frac{t z_4}{t-z_4+1},\frac{t z_5}{t-z_4+1}\biggr)
\end{align}
and
\begin{equation}\label{zdeformation2}
    \Omega_{L{+}1}(z_1,\dots,z_5)=\int_0^\infty  d\log(s+1)\,\,\Omega_{L+\frac12}\biggl( z_1,\frac{z_2}{s+1},z_3,\frac{z_4}{s+1},\frac{z_5}{s+1}\biggr).
\end{equation}

Following the same algorithm in determining last entries of all-loop penta-box integrals \cite{He:2020uxy}, it is straightforward to see that last entries of $\Omega_L(1,4,5,8)$ remain unchanged for $L \geq 2$. 
Recall that with constants $a$ and $b$, last entries of the integral 
\(\int_0^\infty F(t)\otimes (t+b)\ d\log (t+a)\)
are $a$ or $(b-a)$ and 
those of \(\int_0^\infty F(t)\otimes b\ d\log (t+a)\)
are $a$ or $b$. After the first-step integration eq.\eqref{zdeformation1} with $d\log((t+1)/t)$, the five original last entries give six last entries $\{z_1,z_2,z_3,z_4,1-z_4,z_5\}$,
where the new one $1-z_4$ is from the integration 
\[\int_0^\infty F(t)\otimes (t-z_4+1)\,d\log (t).\]
However, deformed $1-z_4$ only contributes $z_4$ as last entry in the second-step integration eq.\eqref{zdeformation2} as well, since after the deformation it only contributes terms like
\[\int_0^\infty F(s)\otimes \frac{s+1-z_4}{s+1}\  d\log (s+1).\]
Therefore by induction, we have proven that last entries of the integral $\Omega_L(1,4,5,8)$ are always $\{z_1,\dots, z_5\}$ for arbitrary $L$.

Using $z$ variables, we also find remarkably simple first-order differential equations:
\begin{equation}\label{PDE1}
 	\Omega_{L+\frac12}=(z_2\partial_{z_2}+z_4\partial_{z_4}+z_5\partial_{z_5})\Omega_{L+1}   
\end{equation}
and 
\begin{equation}\label{PDE2}
	\Omega_{L}=(z_4-1)(z_1\partial_{z_1}+z_4\partial_{z_4}+z_5\partial_{z_5})\Omega_{L+\frac12}.
\end{equation}
For example, consider the deformation of $L+1/2\to L+1$:
\begin{align*}
   \Omega_{L{+}\frac{1}{2}}(z_1,\dots,z_5)&=\int_0^\infty d\log(s+1)\, \Omega_{L}\biggl(z_1,\frac{z_2}{s+1},z_3,\frac{z_4}{s+1}, \frac{z_5}{s+1}\biggr)\\
& =\int_0^{z_5} d\log t\,\, \Omega_{L}\biggl(z_1,\frac{z_2}{z_5}t,z_3,\frac{z_4}{z_5}t, t\biggr),
\end{align*}
its derivative with respect to $z_5$ is 
\[
z_5\partial_{z_5}\Omega_{L{+}\frac{1}{2}}=\Omega_{L}(z_1,\dots,z_5)-\frac{1}{z_5}\int_0^{z_5} d\log t\,\, (z_2t\partial_2+z_4t\partial_4)\Omega_{L} \biggl(z_1,\frac{z_2}{z_5}t,z_3,\frac{z_4}{z_5}t, t\biggr),
\]
where $\partial_2$ and $\partial_4$ denotes partial derivative acting on the second and fourth argument respectively, and then \eqref{PDE1} is given by the following identity
\[
\frac{1}{z_5}(z_2t\partial_2+z_4t\partial_4)\Omega_{L} \biggl(z_1,\frac{z_2}{z_5}t,z_3,\frac{z_4}{z_5}t, t\biggr)=(z_2\partial_{z_2}+z_4\partial_{z_4})\Omega_{L} \biggl(z_1,\frac{z_2}{z_5}t,z_3,\frac{z_4}{z_5}t, t\biggr).
\]
\eqref{PDE2} can be found in a similar way from the deformation of $L\to L+1/2$. With the DE \eqref{PDE1} and the symmetry, it is also easy to see that the last entries of $\Omega_{L}$ for $L\geq 2$ can only be $\{z_1,\dots,z_5\}$ since $\Omega_{L+\frac12}$ is pure for $L\geq 1$.

Finally from the recursion, we can easily impose certain boundary conditions. Since the boundary value of ${\rm d}\log$ form ${\rm d}\log\frac{t+1}t$ diverges at $t=0$ in eq.\eqref{zdeformation1}, deformed function $\Omega_{L}(1,4,5,8)$ should vanish when $t\to 0$, which gives the constraint:
\begin{equation}
\lim_{t\to 0}\Omega_{L}(tz_1,z_2, z_3,tz_4,tz_5)=0.
\end{equation}
We expect that in the space, differential equations \eqref{PDE1}, \eqref{PDE2}, together with boundary conditions, should determine the symbol of  $\Omega_L(1,4,5,8)$ recursively. This will be confirmed up to weight $8$ in the next subsection. 

Finally, as we have mentioned, setting $u_3\to 0$, {\it i.e.} $z_3\to 1$, $\Omega_L(1,4,5,8)$ degenerates to $7$-point ladder integrals $\Omega_L(1,4,5,7)$, which have been computed in \cite{He:2020uxy,He:2021esx} up to $L=4$ easily. We use this colinear limit as a cross check for our result.

\subsection{Locating the integrals and the pattern for algebraic letters}

We have constructed the space with given first entries and adjacency conditions in the section \ref{sec:3.1}, in this subsection we using the conditions above to bootstrap $\Omega_L$ up to $L=4$. As mentioned above, DE and boundary conditions are sufficient for the task, but computationally it is easier if we first impose last-entry conditions and symmetry of the integral. 

To impose DE explicitly, we use the derivative formula of a symbol:
\[
\partial_a (F\otimes w)=F \frac{\partial}{\partial a}\log w.
\]
In practice, the derivative in DE \eqref{PDE1} takes the ansatz $\sum_{i} F_i\otimes z_i$ into
\[
(z_2\partial_{z_2}+z_4\partial_{z_4}+z_5\partial_{z_5})\sum_{i} F_i\otimes z_i=F_2+F_4+F_5.
\]
For the other DE \eqref{PDE2}, one may need to calculate the derivative of letters by
\[
    \frac{\partial W_i}{\partial z_j} = \sum_k \frac{\partial W_i}{\partial f_k}  \frac{\partial f_k}{\partial z_j},
\]
but it is more convenient to first require that the last entries of $F_2+F_4+F_5$ are $\{z_1,z_2,z_3,z_4,1-z_4,z_5\}$ which is proven for $\Omega_{L+1/2}$ in the last subsection, and then the derivative $(z_4-1)(z_1\partial_{z_1}+z_4\partial_{z_4}+z_5\partial_{z_5})$ is trivial.
To impose the symmetry, we rewrite two symmetries 
$z_1\leftrightarrow z_2$ and $z_3\leftrightarrow z_4$ in terms of alphabet. For example, the transformation $z_3\leftrightarrow z_4$ is simply
\begin{align*}
\{
&W_2\leftrightarrow W_5,
W_7\leftrightarrow W_{10},
W_{11}\leftrightarrow W_{16},
W_{13}\leftrightarrow W_{17},
W_{14}\leftrightarrow W_{18},
W_{21}\leftrightarrow W_{23},\\
&
W_{26}\leftrightarrow W_{28},
W_{32}\leftrightarrow W_{37},
W_{36}\leftrightarrow W_{38},
L_3\to\frac{1}{L_1 L_4},
L_4\to \frac{1}{L_1 L_3},
L_5\to \frac{L_5}{L_1 L_3 L_4}\}.
\end{align*}
The other symmetry $z_1\leftrightarrow z_2$ behaves more complicated because under the transformation $z_1\leftrightarrow z_2$, rational letters $\{W_{30},W_{33},W_{35},W_{36},W_{38}\}$ will produce new factors, thus for bootstrapping $\Omega_L$, we only need the other 33 rational letters. It is remarkable that this is exactly the  smaller rational alphabet obtained from the parity-invariant $G_+(4,8)/T$ as mentioned above! It is intriguing that this smaller alphabet is exactly the one that respects the symmetry; for our purpose, it is sufficient to use only these 33 rational letters (plus $5$ algebraic ones).

Before going to higher loops, it is already interesting to re-derive $\Omega_2$ from the bootstrapping strategy. As mentioned in subsection \ref{sec:3.1}, after imposing the last entry condition, there is only one algebraic function left. By imposing DE (with the weight $2$ function being one loop hexagon \eqref{double}) and boundary conditions, we arrive at the unique symbol of $\Omega_2$, which is recorded in ancillary file.

\begin{table}[h]
    \centering
    \begin{tabular}{|c|c|}
       \hline
       conditions &  \# free parameters  \\ \hline 
       weight-6 function space   & 3585 \\ 
       last entry & 257 \\
       symmetry $z_3\leftrightarrow z_4$ & 146  \\
       symmetry $z_1\leftrightarrow z_2$  & 56  \\
       DE & 3  \\
       boundary conditions & 0 \\
       \hline
    \end{tabular} 
    \caption{Number of free parameters left after using constrains on the left column for bootstrapping $\Omega_3$}
    \label{table.1}
\end{table}

We continue to determine $\Omega_3$ in this way: the number of free parameters of the ansatz during the bootstrap of $\Omega_3$ is given in the Table \ref{table.1}. 
Note that the letters in the symbol after imposing the last entries is dramatically reduced, only $29+5$ letters left. These $29$ rational letters behave well under the transformation $z_1\leftrightarrow z_2$. Then imposing the derivative $(z_2\partial_{z_2}+z_4\partial_{z_4}+z_5\partial_{z_5})$, we 
get an ansatz of $\Omega_{2+1/2}$ whose last entries are proven to be $\{z_1,z_2,z_3,z_4,z_5,1-z_4\}$, but here in the ansatz naively we have $9$ extra last entries. It is convenient to just eliminate these ``spurious" last entries, and then apply the second DE and boundary conditions which allow us to immediately determine the symbol of $\Omega_3$.

Next, we want to determine $\Omega_4$. Since the function space (even after using adjacency) is too large at weight 8, we find it useful to directly impose last-entry conditions when constructing the space for weight 7 and above (we also use the alphabet with 33 rational letter which respect the symmetry $z_1\leftrightarrow z_2$). After obtaining the reduced space, it becomes straightforward to apply DE and boundary conditions, which uniquely determine the symbol of $\Omega_4$ (it takes a few hours on a laptop). 

The symbol of $\Omega_3$ has about $8\times 10^4$ terms which is recorded in the ancillary file, while $\Omega_4$ has more than $10^6$ terms which is too lengthy to be recorded. Although the complexity of the result grows fast with the number of loops, we find some hidden simplicity at least for the algebraic part. Just like $\Omega_2$, the part of $\Omega_3$ and $\Omega_4$ containing algebraic letters take a strikingly simple form:
\[
\sum_{i=1}^5 \mathcal{S}(F(2,4,6,8))\otimes L_i \otimes\mathcal{S}(F_i)
\]
where $F_i$ are weight $3$ ($5$) MPL functions with rational letters only, for $\Omega_3$ ($\Omega_4$) respectively. This means that in addition to $L_1, L_2$ in the second entry as part of ${\cal S}(F(2,4,6,8))$, the 5 algebraic letters {\it only} appear on the third entries but not any subsequent ones. This phenomenon was observed at the special $R^{1,1}$ kinematics \cite{He:2021fwf} as well; in $R^{1,1}$ $\Omega_L$ is a rational $A_2$ function  $\Omega_L(v,w)$, and only non-trivial algebraic letters, $L_3, L_4, L_5$ become the ``mixing letter" $v-w$. We have proven that in $R^{1,1}$ kinematics, the part of $\Omega_L(v,w)$ $L$ with mixing letter reads~\cite{He:2021fwf} :
\[
\mathcal{S}(F(2,4,6,8))\otimes(v-w)\otimes\mathcal{S}(\log^{2L-3}(\frac vw))\,.
\]
This indicate that for all $L$, algebraic letters $L_i$ for $i=3,4,5$ can only appear at the third entries (with symbol of $F(2,4,6,8)$ in the first two), but still does not exclude the possibility that simpler algebraic letters, $L_1, L_2$, may appear in subsequent entries. Here we confirm that at least through four loops, no algebraic letters appear beyond the third entry. Recall that $\Omega_L(1,4,5,8)$ is given by a two-fold integral of $\Omega_{L-1}(1,4,5,8)$, and the pattern we observe means that the first three entries of the algebraic part remain unchanged! It is an interesting problem to prove this by carefully analysing rationalization and possible cancellation of spurious square roots in our integration routine. Note that the same phenomenon is expected to hold for three-loop MHV amplitudes, which follow from similar pattern of two-loop NMHV ones via ${\bar Q}$ equations, as observed in~\cite{Caron-Huot:2013vda} in $R^{1,1}$ kinematics.

We also note that the rational alphabet of $\Omega_3$ and $\Omega_4$ does not contain $38$ or even $33$ rational letters but only the first $25$ $\{W_1,\cdots,W_{25}\}$, which can already be found in all non-zero Pl\"ucker coordinates. We believe that other integrals sharing the same kinematics as $\Omega_L(1,4,5,8)$ may contain some of the remaining letters, and we leave it to future work for finding and studying such integrals. 

\section{Conclusion and Discussions}
In this paper, we have conjectured that symbol alphabets for certain classes of DCI Feynman integrals can be determined by truncated cluster algebras purely from their kinematics, which are boundaries of $G_+(4,n)/T$. The main example we study is the two-mass-opposite hexagon kinematics, and our method produces an alphabet of $38$ rational letters and $5$ algebraic ones (as a truncated affine $D_4$ cluster algebra). We construct the space of integrable symbols after imposing physical first-entry conditions and a truncated version of cluster adjacency, which we believe to be universal. When restricting to $\Omega_L(1,4,5,8)$, we derive differential equations and last-entry conditions from our $d\log$ recursion, which allow us to locate its symbol in the space up to weight $8$. We also find a remarkable pattern up to four loops for the appearance of algebraic letters, which begs for some explanations. Since the rationalization is very similar to those required for computing multi-loop amplitudes using $\bar{Q}$ equations, it is tempting to look for similar pattern  for higher-loop $n=8$ amplitudes.

We have only focused on cases where the kinematics can be naturally given in terms of positroid cells of $G_+(4,n)$. If the kinematics for a class of Feynman integrals cannot be labelled by positroids, our method does not directly apply since we do not know the quiver to begin with. For example, currently we do not know any positroid cell for two-mass-hard hexagon kinematics. In~\cite{Chicherin:2020umh}, the alphabet of the latter was conjectured to be a subset of octagon alphabet that are annihilated by first-order differential operators encoding the kinematics. This seems to be a general method when we know the alphabet of $G_+(4,n)/T$, and it would be interesting to study the relation of such subsets to our truncated cluster algebras. For example, if we apply the differential operator to our two-mass-opposite case, we have $33$ of the $38$ rational letters and $5$ algebraic letters. An important difference is that these subsets generally do not correspond to boundaries of $G_+(4,n)/T$ (while we expect our truncated cluster algebras do). We have looked at higher-dimensional cases, {\it e.g.} for one-mass heptagon kinematics with $n=8$, our method gives a co-dimension 2 boundary of $G_+(4,8)/T$ which has $100+1$ facets, where we have 100 $g$-vectors and 1 limit ray (the subset from differential operators of~\cite{Chicherin:2020umh} is smaller). Since the computation for $G_+(4,n)/T$ cluster algebra becomes very difficult beyond $n=8$ (there are recent results for $n=9$ using a subset of all Pl\"ucker coordinates~\cite{Henke:2021avn}), it is crucial to develop both methods  for studying higher-point DCI integrals. It is also an interesting mathematical problem to systematically classify the boundaries of $G_+(4,n)/T$ (see~\cite{Arkani-Hamed:2020cig}) and study their relevance for Feynman integrals. 

Both for amplitudes and integrals in ${\cal N}=4$ SYM, the possibility of studying symbol alphabets pure from from kinematics sounds like magic: despite more and more data supporting such conjectures, we do not have a good understanding of the mechanism. Compared to scattering amplitudes, there might be a better chance to systematically understand why alphabets of certain DCI Feynman integrals are related to such truncated cluster algebra, especially via canonical differential equations~\cite{Henn:2013pwa,Henn:2014qga}. It would be also highly desirable to connect alphabets for these integrals to certain $4k$-dimensional plabic graph of $G_+(k,n)$ as have been studied for amplitudes, which essentially amount to maps from such cells in $G_+(k,n)$ to (boundaries of) $G_+(4,n)/T$. A pressing question is to see if and how more complicated algebraic letters including those containing higher-order roots appear in truncated cluster algebras for corresponding integrals (one could even speculate something ``elliptic" might appear for the ``alphabet" of the kinematics for two-loop $n=10$ double-box integral). Last but not least, cluster algebra structures have been observed for Feynman integrals that are not DCI including those with IR divergence; their alphabet can sometimes be obtained from that in DCI case, {\it e.g.} the pentagon alphabet with one-massive leg can be obtained from sending a dual point to infinity in the two-mass-hard hexagon kinematics~\cite{Chicherin:2020umh}. It would be extremely interesting to find possible truncated cluster algebras for these more general integrals, which should again be directly related to their canonical differential equations. 

\section*{Acknowledgement}
It is a pleasure to thank Nima Arkani-Hamed, Yichao Tang, Yihong Wang, Chi Zhang, Yang Zhang,  Yong Zhang and Peng Zhao for inspiring discussions, correspondence and collaborations on related projects. We would like to thank especially James Drummond and \"Omer G{\"{u}}rdo{\u{g}}an for helpful comments on the first version of the paper. This research is supported in part by National Natural Science Foundation of China under Grant No. 11935013,11947301, 12047502,12047503.
\bibliographystyle{utphys}
\bibliography{main}
\end{document}